\documentclass[aps,prb,twocolumn,showpacs,superscriptaddress,floatfix,nofootinbib]{revtex4}
\usepackage{graphicx}
\usepackage{amssymb}
\usepackage{amsmath}
\usepackage[english]{babel}
\def\bfig{\begin{equation}gin{figure}}
\def\efig{\end{figure}}

\newcommand{\Rangle}{\mbox{{$\rangle\kern -2pt\rangle$}}}

\newcommand{\be}{\begin{equation}}
\newcommand{\ee}{\end{equation}}
\newcommand{\bea}{\begin{eqnarray}}
\newcommand{\eea}{\end{eqnarray}}
\newcommand{\bmat}{\begin{pmatrix}}
\newcommand{\emat}{\end{pmatrix}}
\newcommand{\bs}{\begin{split}}
\newcommand{\es}{\end{split}}

\newcommand{\f}{\frac}

\newcommand{\kaava}[1]{(\ref{#1})}

\newcommand{\corr}[1]{\left<#1\right>}

\newcommand{\im}{i}

\newcommand{\doo}{\partial}

\newcommand{\abs}[1]{\left\vert#1\right\vert}
\newcommand{\de}{\mathrm{d}}

\newcommand{\real}{\mathbb{R}}
\newcommand{\complex}{\mathbb{C}}
\newcommand{\rational}{\mathbb{Q}}
\newcommand{\integers}{\mathbb{Z}}

\newcommand{\set}[1]{\{#1\}}
\newcommand{\bset}[1]{\left \{#1\right \}}

\newcommand{\halfspace}{\mathbb{H}}
\newcommand{\chspace}{\overline{\mathbb{H}}}

\renewcommand{\Im}[1]{\textrm{Im }#1}
\renewcommand{\Re}[1]{\textrm{Re }#1}

\newenvironment{matriisi}{\left(\begin{matrix}}{\end{matrix}\right)}

\begin{document}

\title{The quantum Hall curve}
\author{J. Nissinen}
\author{C.A.  L\"utken}
\affiliation{Theory group, Department of Physics, University of Oslo, N-0316 Oslo, Norway}

\date{\today}

\begin{abstract}
We show how the modular symmetries that have been found to be consistent 
with most available scaling data from quantum Hall systems\cite{LR1, LN2}, 
derive from a rigid family of algebraic curves  of the elliptic type.  
The complicated special functions needed to describe scaling data arise
in a simple and transparent way from the group theory and geometry of 
these \emph{quantum Hall curves}. 
The  renormalization-group potential therefore emerges naturally in a geometric context that 
complements the phenomenology found in our companion paper\cite{LN2}.  
We show how the algebraic geometry of elliptic curves is an efficient
way to analyze specific scaling data, extract the modular symmetries of the transport 
coefficients, and use this information to fit the given system into the
one-dimensional (real) family of curves that may model all universal properties 
of quantum Hall systems.
\end{abstract}
\maketitle

\section{Introduction}

Since the discovery of the quantum Hall effect (QHE) three decades ago, 
a huge amount of experimental data have been collected, but a comprehensive
theory that accounts for all universal properties of these systems is still missing.

The situation is somewhat reminiscent of particle physics before the discovery of 
gauge symmetries.   The zoo of  ``elementary" particles appeared to be organized 
in simple patterns that turned out to be weight diagrams of some approximate 
global \emph{unitary} symmetries.  The origin of these phenomenological symmetries
in the exact underlying gauge theory was only understood much later, but they played 
an important role in the development of the standard model.

It is still not known how to prove that these symmetries emerge from quantum chromodynamics,
but this gap in our understanding of the mathematics of non-abelian gauge theories has not impeded the progress of particle physics.  An effective field theory (EFT) known as 
``the chiral model" can be constructed using the observed low energy degrees of freedom
 (hadrons) and their approximate global symmetries as input, 
 and this is sufficient to model the low energy dynamics.   
In order for particle physics to evolve from 
particle taxonomy to deep underlying principles it was essential to acquire new 
mathematical tools, including some group theory and geometry.

Similarly, almost all the scaling data in quantum Hall systems appears to be organized 
in simple patterns that are phase diagrams of some approximate global \emph{modular}
symmetries \cite{LR1,LN2}. Their origin must be in the exact underlying gauge theory 
(quantum electrodynamics in disordered media), but it is not known how to prove 
that they emerge as effective symmetries at the low energies used in transport experiments.
So far efforts to construct an EFT start by postulating some low energy degrees of freedom
(see e.g. Refs.\,\onlinecite{Pruisken, ASbook, KLZ, BurgessDolan}).

It does not seem implausible that modular symmetries 
will aid in the construction of effective field theories that can model
the emergent behavior observed in two-dimensional electron liquids,
but in order to progress beyond the taxonomy of scaling diagrams it is necessary 
to acquire new mathematical tools that are not in general use today in 
condensed matter physics.  Our primary purpose here is therefore to 
explore the mathematical structure underlying the observed modular symmetries,
and to show how the data are neatly encoded in the geometry of an algebraic curve
that we shall call \emph{the quantum Hall curve}.

These algebraic curves are not without precedent in physics.
As is often the case, Nature appears to recycle a good idea, 
and similar structures appear in integrable models of statistical mechanics.
It was Onsager who realized the utility of elliptic functions in this context,
which flourished in the Baxter model and its more recent generalizations
using algebraic curves (see e.g. Refs.\,\onlinecite{Baxter, Maillard}).   
They are also the centerpiece of the Seiberg-Witten 
theory of certain supersymmetric models, where the symmetries are sufficiently rigid to allow
the construction of the full low-energy effective action from a certain elliptic curve \cite{SW}. 
In both cases what is required is to have two non-commuting discrete symmetries in parameter space that are combinations of translations and duality transformations.
Translation symmetries are ubiquitous in physics, while
dualities are not uncommon, being relatives of Kramers-Wannier duality
 in statistical mechanics, and of  electro-magnetic duality in field theory. 

In order to establish notation we recall that a quantum field theory is 
defined by a generating function constructed by summing over all field configurations 
$\varphi$ weighted by the piece of the action $S(\varphi; \tau)$ available to the 
physical system in the experiment under consideration. 
In other words, the coupling constants $\tau = (\tau_1,\tau_2,\dots)$ 
span the relevant part $\mathcal M$ of the full (infinite dimensional) 
parameter or \emph{moduli} space of all conceivable field theories.  
More precisely, in the correct renormalization group (RG) vernacular, 
$\mathcal M$ is the finite dimensional subspace spanned by 
relevant and marginal operators.

Restricting attention to this finite dimensional moduli space is necessary,
in order to do physics, but it is not sufficient.  Almost all field configurations 
cost too much energy, and the effective field theory should retain only those that are
accessible at low energy.  The conventional (``bottom-up") approach 
is to focus first on the effective action, trying to extract some of its 
properties from the microscopic theory, and subsequently derive some of the 
properties of $\mathcal M$ by calculation.  But in quantum Hall experiments 
it is $\mathcal M$ that is probed directly, so our phenomenological 
(``top-down") approach is the reverse, as we now summarize in the briefest 
possible terms. 

The symmetries have been extensively discussed in a companion paper
\cite{LN2}, where we analyze available scaling data and show that almost all data 
fit into a one-parameter family of RG potentials with modular symmetry.
The theory of modular groups will therefore not be our main focus here, but rather 
the underlying geometry to which they are naturally associated.
Information on modular groups and their representations has been 
collected in the Appendix for easy reference. 

Since it may not be completely obvious why symmetries, especially intricate 
modular symmetries, are inextricably entangled with geometry, 
it may be helpful to recall Felix  Klein's ``Erlangen Program" \cite{Klein}.
He proposed that abstract symmetries  (group theory) organizes geometry, 
and that projective geometry is the unifying framework.
Again the situation is somewhat reminiscent of the inception of gauge theories.
Quantum electrodynamics was initially a rather awkward construction based on the 
obscure idea called ``minimal substitution", which eventually matured into 
$U(1)$ gauge invariance.  It is possible to generalize this symmetry to $SU(n)$ 
by ``brute force", but this is greatly facilitated by using principal fiber bundles,
which is the ``Kleinian geometry" where gauge symmetries act naturally.
Similarly, the Kleinian geometry to which modular symmetries belong is 
the complex projective geometry of algebraic curves, and more specifically,
the hyperbolic geometry of their moduli spaces.

Consider first the spin polarized QHE. 
From plateaux and scaling data we first infer that the moduli space 
is the modular curve $\mathcal M_0(2)$ with $\Gamma_0(2)$ symmetry,  
and that the RG $\beta$-function is a holomorphic vector field on this 
curve~\cite{LR1}.
Because the RG flow appears to respect both the symmetry and the complex
structure of $\mathcal M_0(2)$ this is a powerful result:
the identification $\mathcal M \simeq \mathcal M_0(2)$ seems to
fix almost all global universal data, and we are led to consider the 
family $\mathcal E_0(2)$ of elliptic curves whose moduli space is
$\mathcal M_0(2)$.

So far the geometry of these elliptic curves, which are ``rigidified" 
(also called ``framed", ``enhanced" or ``decorated" in the mathematical literature) 
by certain torsion data, have received little attention in physics, apart from their relation to spin structures on the torus.  
One reason is that so much of the physically relevant data about
universality classes is encoded directly in the topology and geometry of the moduli space.
In the polarized case, $\mathcal M_0(2)$ is so rigid that it predicts the exact 
location of all quantum critical points, as well as the precise shape of all RG flow lines. 
Remarkably, this is in agreement with a host of experiments.  
One of the coldest experiments to date \cite{Tsui6}, where the effective field theory 
is expected to be most accurate, has verified the modular prediction from 1992 
at the \emph{per mille} level \cite{LR1}.

There is no obvious physical reason why the phenomenological approach should be
restricted to the polarized system, provided we are willing to consider different symmetries, which 
usually will be smaller than $\Gamma_0(2)$ (but never larger).  The fully polarized case is 
continuously connected to the unpolarized system by tuning one or more suitable control parameters, at least in principle. This raises the question how these ``symmetry transitions", 
which appear to be discontinuous, can come about.  
We address this question by studying a family of algebraic curves, which 
gives rise to a surprisingly simple family of RG potentials $\Phi_a$ that interpolates 
between the points of enhanced symmetry.

Another compelling reason to study these curves is to understand the origin and 
properties of holomorphic vector fields on $\mathcal M$, 
which include the $\beta$-functions whose integral curves are the RG flow lines in
these models.  In maximally symmetric cases they are essentially unique,
and less symmetric cases at level 2 are only slightly less constrained.
The additional freedom is sufficient, but just barely so, to account for unpolarized data.
 This is one example of a remarkable confluence of quantum Hall physics
and modular mathematics, which is best appreciated in a geometric setting.

The absence of a derivation of these symmetries from a microscopic model of 
charge transport makes it difficult to compare our phenomenological analysis with
more conventional models.   This applies to all three regimes of interest: 
the infrared (IR) plateaux domain where Laughlinesque wave functions and Chern-Simons type topological theories capture the
relevant physics; the ultraviolet (UV) perturbative domain 
where non-linear sigma models \cite{Pruisken} describe 
the relevant modes (for a current review see Ref.\,\onlinecite{ASbook}); 
and the scaling domain where a conformal field theory is expected to appear,
albeit a rather subtle one.  The UV/IR  limits are at the boundary of $\mathcal M$, 
infinitely far away from the scaling region in any natural metric on moduli space.  
This is, perhaps, where one would expect our effective field theory
to break down, and be replaced by other models better adapted to the UV/IR limits.
Indeed, using the algebraic approach advocated here, we will show that
quantum Hall curves degenerate to singular geometries in these limits.
Our approach has been tailored for the quantum critical regions controlled by 
quantum critical points deep inside moduli space.  
A deep understanding of the geometry of the quantum Hall curves at 
these points may aid in the construction of the conformal quantum critical model.

The next section is a primer on elliptic curves that provides the theoretical 
framework and mathematical tools required for the experimental analysis,
presented in a companion paper \cite{LN2}.  One of the main goals in this section 
is to clarify the algebraic and geometric origin of the special (elliptic and modular) functions that 
we use extensively.    We also explain the geometric origin of families of curves 
that have modular symmetries smaller than
the full modular group ${\rm PSL}(2,\mathbb Z)$, and we discuss all congruence 
sub-groups at level two. This appears to be sufficient for the analysis of 
unpolarized quantum Hall data \cite{LN2}.

In Sec.\,III we turn our attention to a more detailed discussion of 
the modular curve $\mathcal M$.   
Modular functions that live on the modular curve $\mathcal M$
arise from elliptic functions that live on the elliptic curve $\mathcal E$,
encountered in the previous section.

This equips us with the technology to explain in Sect.\,IV how scaling experiments 
inform us of local and global properties of the renormalization group, 
and how to reconcile this (semi-)group with modular symmetries.
The resulting $\beta$-functions are rather complex holomorphic modular 
vector fields with the remarkable property that they generate gradient flows.   
This is both physically interesting, since RG potentials are 
expected to exist on general grounds, and also a great mathematical simplification, 
since we need only consider scalar fields on $\mathcal M$.  
Combined with some reasonable physical assumptions about RG flows in general, 
this finally leads us to propose a simple one parameter family of curves with 
RG potentials $\Phi_a$  invariant under level two modular symmetries.  
Our data analysis \cite{LN2} indicates that this potential is consistent 
with almost all available scaling data, for $a\in\mathbb R$.

\section{Elliptic curves}

\begin{figure*}[t]  
\begin{center}
\includegraphics[scale = .42]{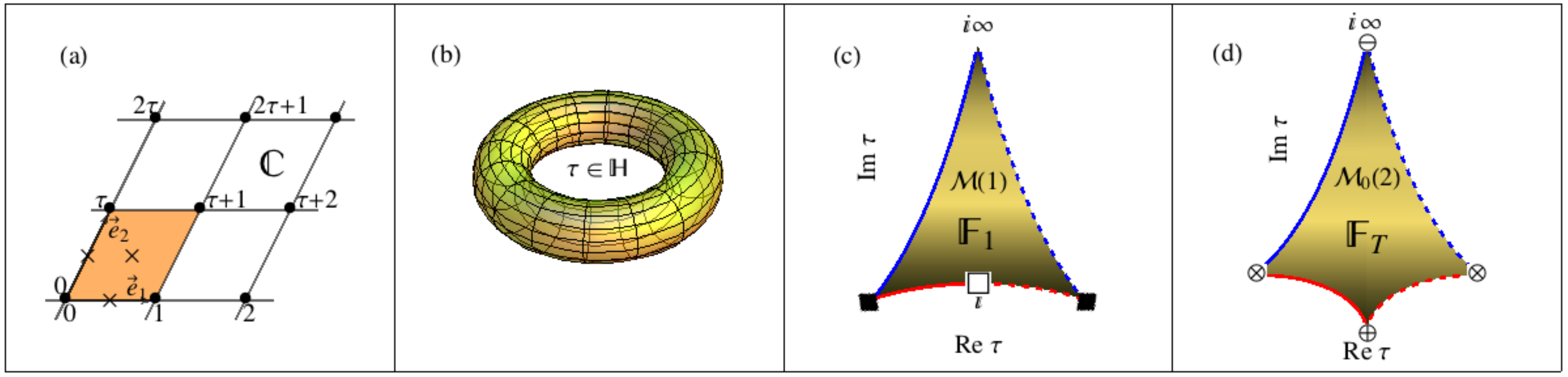}
\end{center}
\caption[LN1_Fig1]{(Color online) An \emph{elliptic curve} is both (a) a lattice $\Lambda_\tau$
(crosses are the 2-torsion points mod $\Lambda_\tau$), and 
(b) a genus $g = 1$ Riemann surface. 
(c) In-equivalent elliptic curves are uniquely parametrized by the points of the
hyperbolic triangle $\mathbb F_1$, which is the interior of the \emph{modular curve} 
$\mathcal M(1)$ of genus $g = 0$.
(d) The polarized quantum Hall curve $\mathcal E_0(2)$ has symmetry 
$\Gamma_0(2)$ with fundamental domain  $\mathbb F_{\rm T}$, and is 
parametrized by the modular curve $\mathcal M_0(2)$.}
\label{fig:LN1_Fig1}
\end{figure*}

The modular symmetry observed in the QHE is intimately related to a special type of 
Riemann surface, which belongs to a family of so-called \emph{elliptic curves}.
In the jargon of nineteenth century geometry,  
this curve is a ``1-pointed genus-1 curve with 2-torsion".  The unfortunate 
use of both ``curve" and ``surface" for the same object derives from 
the convention in algebraic geometry that curves by default are defined over 
$\mathbb C$, while Riemann surfaces are over $\mathbb R^{2}$. 
So a ``curve" has one complex dimension but two real dimensions.  
The genus $g$ is the number of holes in this real 2-dimensional surface.  
The r\^ole of one (origin) or more (torsion) privileged points is explained below. 
In order to understand the properties of these models, in particular the 
existence and uniqueness of the holomorphic modular vector fields used
extensively in our analysis of the scaling data, we need a primer on 
this geometric structure, which we now supply.

\subsection{Topology of tori}

The flat 2-dimensional torus $T^2$ is by definition the surface obtained from $\real^{2}$ by the identifications
\be
(x,y)\simeq (x+1,y)\simeq (x,y+1).
\ee
In other words, topologically a torus is just the product of two circles, 
$T^2 \simeq S^1\times S^1$. 

Any torus can be mapped to any other by a continuous change of real coordinates, 
so unless we endow them with some additional structure they are rather featureless.   
Demanding smoothness does not refine this classification: 
the genus $g$ is the only invariant under homeomorphisms or diffeomorphisms.
In order to model the QHE quite a lot of ``decoration" with extra data is needed:
the additional data are called \emph{complex structure} and \emph{torsion}.

We first explain what a complex structure is.  
Two tori are only equivalent, i.e., diffeomorphic,
as complex manifolds if there is a holomorphic (complex analytic)
coordinate transformation between them, and this is often not possible.
Two tori that are distinct in this sense are said to have different complex structures. 
One way to parametrize this structure is by introducing the complex coordinate $z:=x+\tau y$
($\Im \tau\neq 0$) on the torus. 
The periodic identifications dictated by the topology are now
\be
z \simeq z+1 \simeq z+\tau. 
\ee
In other words, we can represent the torus as the complex plane ``rolled up" 
in both directions, i.e., $T^2_\tau = {\mathbb C}/\Lambda_\tau$, 
where $\Lambda_\tau =  {\mathbb Z} + {\mathbb Z}\tau$
is the skewed lattice in $\mathbb C$ shown in Fig.\,\ref{fig:LN1_Fig1}\,(a)
with basis vectors $\vec{e}_1 = (1,0)$ and $\vec{e}_2 = (0,\tau)$.
By choosing an ordering of the basis we can limit $\Im \tau>0$.  Notice that 
we have already used the ``small'' complex diffeomorphisms of the torus: 
a translation to fix the origin of the lattice, and a scaling to normalize the first basis vector.
It is obvious that the geometry and topology of a genus-1 curve is that of a torus,
compare Fig.\,\ref{fig:LN1_Fig1}\,(b). Once an origin of the lattice has been 
singled out, the torus is called \emph{an elliptic curve}. 

It is clear that $\tau$ labels all possible complex 
structures, but in a highly redundant way.  
We can restrict $\tau$  without loss of generality to the upper half of the complex plane, 
\begin{equation*}
\tau\in\mathbb H := \{\tau\in\mathbb C\vert \Im \tau >0 \} \;,
\end{equation*}
but this still leaves an infinite number of ``copies" $T^2_{\tau^\prime}$ of each complex 
torus $T^2_\tau$.  In fact, $\tau^\prime$ represents the same lattice, i.e., the same complex structure, as 
$\tau$ iff $\tau^\prime = \gamma(\tau)$, where $\gamma$ is a M\"obius transformation 
taken from the modular group $\Gamma(1) = {\rm PSL}(2,\mathbb Z)$. These are the ``large'' diffeomorphisms of the torus. How this comes about in terms of equations is detailed below, when considering elliptic curves. There is one last discrete diffeomorphism of the torus with the complex structure $\tau$ in the upper half plane $\mathbb{H}$, which corresponds to changing the orientation of $\Lambda_{\tau}$, acting as $\tau \mapsto -\overline{\tau}$ and not a modular transformation. Instead, it is an outer automorphism of $\Gamma(1)$, and arises in the QHE as the particle-hole conjugation. Since this is simply a reflection in the imaginary axis of
 $\mathbb{H}$, its consequences are rather simple and we will focus on the modular transformations.

The modular group partitions $\mathbb H$ into equivalence classes, and a region of $\mathbb H$ that 
represents each possible complex structure precisely once is called a fundamental 
domain $\mathbb F_1$ of $\Gamma(1)$.  The simplest choice of this domain is to take
$\tau$ above the unit circle in the strip $\vert\Re\tau\vert \leq 1/2$,
\begin{equation*}
\mathbb F_1(\tau) = \{\tau\in\mathbb H\vert  -1/2\leq\Re\tau < 1/2 \;,
\vert\tau\vert \geq 1\}\;.  
\end{equation*}
Identifying all points at infinity, we can think of the fundamental domain 
as the hyperbolic triangle shown in Fig.\,\ref{fig:LN1_Fig1}\,(c).
Identifying points on the boundary of $\mathbb F_1$ that are connected by modular
transformations gives the moduli space $\mathcal M(1)$.  Topologically it is also a curve,
and in fact an algebraic curve of genus-0, i.e., the Riemann sphere, which is called 
\emph{the modular curve} for the full modular group.

Adding more structure breaks the symmetry to a subgroup $\Gamma\subset\Gamma(1)$,
splitting up the equivalence classes further so that a larger
fundamental domain is needed in order to label them all. 
For example, in the spin-polarized quantum Hall case the modular symmetry is 
$\Gamma_0(2)$, which has the fundamental domain
\begin{equation*}
\mathbb F_{\rm T}(\tau) = \{\tau\in\mathbb H\vert -1/2\leq\Re\tau < 1/2\;, 
\vert\tau\pm 1/2\vert \geq 1/2\}\;.
\end{equation*}
Identifying all points at infinity, we can think of the fundamental domain 
as the hyperbolic polygon shown in Fig.\,\ref{fig:LN1_Fig1}\,(d).
Identifying points on the boundary of $\mathbb F_{\rm T}$ that are connected by
modular transformations taken from $\Gamma_0(2)$ gives the level 2 moduli space 
$\mathcal M_0(2)$.  
Topologically it is also a curve, and again an algebraic curve of genus-0, 
which is called the modular curve for the congruence subgroup 
$\Gamma_0(2) \subset \Gamma(1)$, sometimes abbreviated to ``submodular curve".
We shall soon meet other parameterizations of the complex structure of 
elliptic curves.

\subsection{Algebraic curves}

In order to explain the additional structure (torsion) needed to model the QHE,  
we turn to the algebraic representation of elliptic curves. 
It is far from obvious that an \emph{algebraic curve}, 
i.e., a dimension one subset of a complex projective space, 
has anything to do with two-dimensional tori (elliptic curves).

The simplest algebraic presentation of a torus is as a ``planar cubic". 
The complex projective ``plane" $\mathbb C\mathbb P^2$ has four real dimensions, 
which is reduced to two real dimensions by any sufficiently general (``complete") 
algebraic (polynomial) constraint, since the polynomial is also defined over $\mathbb C$. 
The hypersurface defined by the vanishing locus of the polynomial is therefore a 
Riemann surface,  i.e.,  a complex curve.   If the polynomial has degree three 
then it is an elliptic curve.

There are infinitely many other ways to present elliptic curves, 
by algebraic constraints in ambient projective spaces of any dimension,
but so-called ``complete intersections" are especially simple and therefore especially useful.  
There are only two:  planar cubics and the intersection of two quadrics in 
$\mathbb C\mathbb P^3$.
The black skewed squares in Fig.\,\ref{fig:LN1_Fig1}\,(c) are pictograms of tori with these 
shapes, which are simply related to planar cubics. 
The white square icon at $i = \sqrt{-1}$ in Fig.\,\ref{fig:LN1_Fig1}\,(c) is a pictogram of 
a torus with this shape, which is simply related to the intersection of two quadrics.

Independent of any specific presentation, an elliptic curve has two remarkable properties.
First, it is always isomorphic to a torus, and any (marked) torus is isomorphic to an elliptic curve.
This means that there is always a (highly transcendental) correspondence between the 
coefficients of the defining polynomials and the complex number $\tau$ labeling tori. 

Second, unlike all other curves ($g\neq 1$), a genus-1 curve can be endowed with 
a group structure.  This is the source of its remarkable arithmetic properties that 
even after two centuries of intense scrutiny remains an active field of research.
More precisely, a genus-1 curve is always an \emph{abelian group} provided one point 
(in the following construction, the point $O$ at infinity) is chosen as the zero element for the ``chord-tangent" group law, 
which to any two points $P$ and $Q$ on the curve gives a third point $P+Q$.   The freedom 
in choosing the identity of this composition rule corresponds to the freedom to translate the toroidal lattice $\Lambda_\tau$ to choose any point to serve as the origin, since the abelian group law is equivalent to simply addition in $\complex/\Lambda_{\tau}$.
By definition an elliptic curve comes equipped with this point, but a topological torus without a complex structure does not.

\subsection{Planar cubics}

By a suitable change of coordinates we can always bring a cubic polynomial in 
$\mathbb C\mathbb P^2$ to the form $y^2 = \mathcal{E}(x)$, where $\mathcal{E}(x)$ is a cubic 
polynomial in $x$, and $x = X/Z$ and $y=Y/Z$ are inhomogeneous coordinates 
obtained from the homogeneous coordinates $X,Y$ and $Z$ of 
$\mathbb C\mathbb P^2$ on a chart with a constant $Z\neq 0$. 
In order to see this we first observe that a general cubic constraint with a marked point $O$ can be written \cite{Silverman}
\begin{equation*}
ZY^{2} + a_{1}XYZ + a_{3}YZ^{2} = X^{3} + a_{2}X^{2}Z+a_{4}XZ^{2}+a_{6}Z^{3} 
\end{equation*}
after a suitable rescaling. This curve has the special point $O$ at infinity given by 
$X = Z = 0$ and $Y\neq 0 $.  Since  $Z \neq 0$ elsewhere on the curve, 
the inhomogeneous form of this equation is
\begin{equation}
\mathcal{E}: y^{2} + a_{1}xy + a_{3}y = x^{3}+a_{2}x^{2}+a_{4}x+a_{6} \;.
\label{eq:cubic}
\end{equation}

To what extent is this equation unique, i.e., how does the elliptic curve $\mathcal{E}$ 
(with $O$ fixed at infinity) depend on the constants $a_{1}$, $a_{2}$, $a_{3}$, $a_{4}$
 and $a_{6}$? An \emph{admissible} change of variables,
i.e., a coordinate transformation which preserves the above form and retains the special 
point $O$ at infinity, is given by
\begin{equation}
x = u^2  x' + r\;,\quad y = u^3 y' + s u^2 x' + t \;,\quad u\neq 0\;.
\label{eq:admissible}
\end{equation}
The composition of two admissible changes of variables is again admissible.
Since we work over $\complex$, we can use the admissible coordinate transformation 
$y\mapsto (y - a_1 x - a_3)/2$ to complete the square, bringing the equation to the 
much simpler form
\begin{equation*}
y^2 = 4x^3 + b_2 x^2 + b_4 x + b_6\;.
\end{equation*}
Another admissible coordinate transformation given by  
$(x,y)\mapsto (x - b_2/12, y)$ eliminates the quadratic term,
leaving us with a deceptively simple looking two-parameter family of cubics:
\begin{equation*}
\mathcal{E}_{A,B}: y^2  =  4x^3 + A x + B\;.
\end{equation*}
The only change of variables that preserves this form is 
\begin{equation}
(x,y)\mapsto (u^2 x, u^3 y) \;,
\end{equation}
which suggests that the invariant combination 
\begin{equation*}
j \propto \f{A^{3}}{A^{3}+27B^{2}}
\end{equation*}
may be useful.  Whatever the transcendental connection is between the complex structure 
$\tau$ of the torus and the algebraic complex coefficients $A$ and $B$, 
this so-called ``$j$-invariant" is well defined as long as $A^3 + 27 B^2 \neq 0$.

Retracing the steps back to the general cubic in Eq.\,(\ref{eq:cubic}), 
it is a long but straightforward calculation to show that the $j$-invariant is indeed preserved, 
i.e., that $j^\prime = j$ under any admissible change of variables given by
Eq.\,(\ref{eq:admissible}). Moreover, all changes of variables between the different forms of the cubics are admissible, so one can work with any of the above forms for a given curve. These presentations of the cubic are collectively known as 
``the Weierstrass form".

As already mentioned, an elliptic curve $\mathcal E$ is also defined by a lattice 
$\Lambda(\omega_1, \omega_2) = \integers\omega_1+\integers\omega_2$ in $\complex$:
\begin{eqnarray*}
\mathcal E  &\cong&   \complex/\Lambda(\omega_1, \omega_2) \\
&=& \left\{z \simeq z+m\omega_{1}+n\omega_{2}\vert z\in\complex ;\, m,n\in\integers\right\}, 
\end{eqnarray*}
 where $\omega_{1}/\omega_{2}\notin \real$, and we can take $\Im \omega_{1}/\omega_{2}>0$
 without loss of generality. But two different lattices can give rise to an equivalent Weierstrass form under admissible change of variables.
 
 First, two bases $(\omega_1, \omega_2)$ and $(\omega'_1, \omega'_2)$ 
 span the same lattice iff
 \begin{equation*}
\begin{pmatrix} 
\omega_1\\ 
\omega_1
\end{pmatrix}  = 
\begin{pmatrix} 
a&b\\
c&d
\end{pmatrix}
\begin{pmatrix}
\omega'_1\\
\omega'_2 
\end{pmatrix} \;,
\qquad \gamma = 
\begin{pmatrix} 
a&b\\
c&d
\end{pmatrix}  \in \Gamma(1) \;.
\end{equation*}
Two lattices $\Lambda$ and $\Lambda^\prime$ will be considered equivalent if
$\Lambda' = \alpha\Lambda$ with  $\alpha\in\complex^{\times}$
(we write this as $\Lambda\sim\Lambda'$), since it will transpire that the 
corresponding elliptic curves are isomorphic, i.e., have the same Weierstrass form. We have seen how to use this freedom to determine equivalent complex structures $\tau$ of the torus, where they arise as complex diffeomorphisms. 
That is, we can always rescale with $\omega_2$, so that 
\begin{equation*}
\Lambda(\omega_1, \omega_2) \sim \Lambda(1, \tau = \omega_1/\omega_2) =
\Lambda_{\tau} = \integers + \integers\tau \;, 
\end{equation*}
where $\Im \tau > 0$ without loss of generality.
From this it follows that 
\begin{equation*}
\Lambda_{\tau_1}\sim\Lambda_{\tau_2} \iff 
\tau_2 = \gamma(\tau_1) =  \frac{a\tau_1 + b}{c\tau_1 + d}
\end{equation*}
with $\gamma\in \Gamma(1)$ and $\tau_1, \tau_2\in\mathbb H$. 
To see this, notice that 
$\Lambda_{\tau_1} = \alpha \Lambda_{\tau_2} = \integers\alpha+\integers\alpha\tau_2$, 
which means that $\alpha\cdot 1= c \tau_1+ d$ and $\alpha \tau_2 = a \tau_1 + b$, 
so $\tau_2 = \gamma(\tau_1)$ and $\alpha  = c\tau_2 + d$. 
The converse follows by choosing $\alpha = c\tau_2 + d$, 
and using that the resulting bases are then related by a $\Gamma(1)$ transformation.

An explicit form for the transcendental transformation between the complex structure 
parameter $\tau$ (the lattice $\Lambda_\tau$) and the elliptic curve $\mathcal{E}$ 
in Weierstrass form, is given by the Weierstrass $\wp$-function
\be
\wp_{\Lambda}(z) = \f{1}{z^{2}} +
\sum_{\omega \in \Lambda/\set{0}}\left(\f{1}{(z-\omega)^{2}}-\f{1}{\omega^{2}}\right)\;,
\ee
which is invariant to lattice transformations, 
$\wp_{\Lambda}(z+\Lambda) =\wp_{\Lambda}(z)$.
The $\wp$-function satisfies the differential equation
\begin{equation}
(\wp_{\Lambda}')^2  = 4 \wp_{\Lambda}^3 - g_2(\Lambda) \wp - g_3(\Lambda)\;.
\label{eq:Weierstrass}
\end{equation}
With $x  = \wp$ and $y = \wp'$ we therefore have an explicit embedding of the torus 
$\mathbb C/\Lambda_\tau$ into the complex projective plane, provided we can always find $\Lambda$ so that $g_{2}(\Lambda)=-A$ and $g_{3}(\Lambda)=-B$.

Two curves will have identical $j$-invariants if $\Lambda' = \alpha\Lambda$, since  $\wp_{\alpha\Lambda}(\alpha z) = \alpha^{-2}\wp_{\Lambda}(z)$ so $(x,y)\mapsto (\alpha^{-2}x,\alpha^{-3}y)$, which is an admissible change of variables with $j=j'$. We have already seen that this means that $\Lambda_{\tau'} = (c\tau+d)\Lambda_{\tau}$, so we immediately have the properties
\bea
g_{2}\left(\f{a\tau+b}{c\tau+d}\right) &=& (c\tau+d)^{4}g_{2}(\tau)\\
g_{3}\left(\f{a\tau+b}{c\tau+d}\right) &=& (c\tau+d)^{6}g_{3}(\tau)\\
j\left(\f{a\tau+b}{c\tau+d}\right) &=& j(\tau) \;.
\eea
The coefficients $g_{2}(\tau):=g_2(\Lambda_{\tau})$ and $g_{3}(\tau):= g_3(\Lambda_{\tau})$ 
are modular forms of weight $4$ and $6$, respectively,  and they generate the ring of all modular forms for $\Gamma(1)$ to be discussed below.  Using the definition of $j(\tau)$ and the 
modular properties of $j$, $g_2$  and  $g_3$, one can complete the equivalence of elliptic curves $\mathcal{E}_{\tau}$ and lattices $\Lambda_{\tau}$ by showing that there always exist a $\Lambda_{\tau}$ such that $g_{2}(\tau)=-A$ and $g_{3}(\tau)=-B$. \cite{Silverman}

The important point is that we see how the elliptic and modular functions parametrize the 
elliptic curve, i.e., the torus with complex structure and origin, 
and how their transformation properties respect the invariances of these objects.

In summary, the following objects are equivalent: 
\begin{itemize}
\item[\emph{i)}] Tori with complex structure $\tau$.
\item[\emph{ii)}] Lattices $\set{\Lambda_{\tau} = \integers+\integers\tau \vert \tau\in\halfspace/\Gamma(1)}$ in $\complex$.
\item[\emph{iii)}] Cubic equations $y^{2}=4x^{2}+Ax +B$ in $\complex\mathbb{P}^{2}$ with 
$j(\tau) \propto A^3 / (A^3+27B^2)$.
\end{itemize}

The modular curve $\halfspace/\Gamma(1)$ is a moduli space of elliptic curves,
i.e., it is a space that parametrizes equivalence classes of curves that are isomorphic
under modular transformation of the lattice and admissible coordinate transformations 
that preserve the origin.  The modular invariant function $j(\tau)$ takes 
different values for different (equivalence classes of) elliptic curves.

\begin{figure}[t]
\begin{center}
\includegraphics[scale = .6]{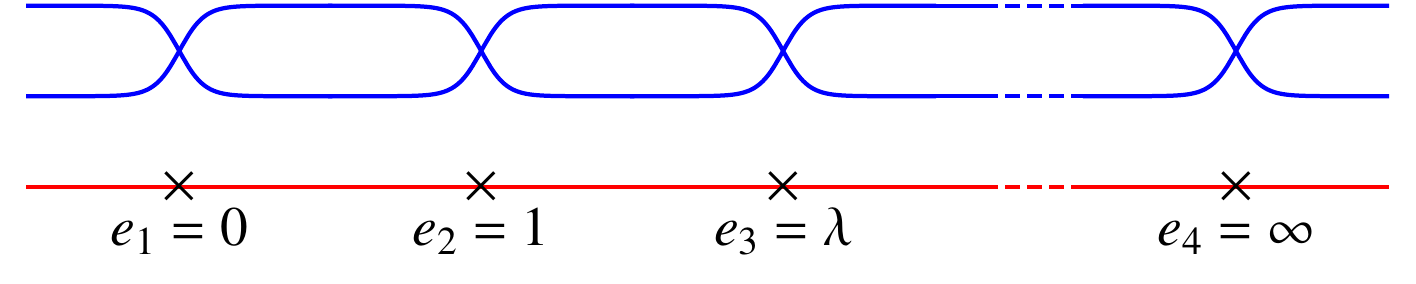}
\end{center}
\caption[Legendre_curve2]{(Color online) Branched covering of the elliptic curve in $\complex\mathbb{P}^{2}$ over $\complex\mathbb{P}^{1}$, with branch points $e_{1}$, $e_{2}$, $e_{3}$
 and $e_{4}$ (adapted from a diagram on p.186 in Ref.\,\onlinecite{DH}).}
\label{fig:LN1_Fig2}
\end{figure}

In the following we shall also discuss subgroups 
$\Gamma(2)\subseteq \Gamma\subset \Gamma(1)$ of the modular group. 
The (sub-)modular curves $\halfspace/\Gamma$ are 
moduli spaces of elliptic curves with additional structure, and
these will have their own special functions. 
This structure is intimately related to 
the inverse of the $\wp$-map discussed above, taking a torus to an elliptic  curve.  
It can be shown \cite{DH} that elliptic curves are mapped to tori by the complex 
valued hypergeometric function $F(1/2,1/2,1;\lambda)$.  
This is in general a two-to-one map, as illustrated in Fig.\,\ref{fig:LN1_Fig2}, 
except at the branch points $e_i$, which have special group theoretical properties
called ``2-torsion".

\subsection{Torsion points}

The simplest way to understand the abelian group law, and the special points on $\mathcal{E}_{\tau}$ 
related to this, is to consider addition modulo $\Lambda_{\tau}$ on $\complex$. 
The embedding $\mathcal{E}_{\tau}\hookrightarrow \complex\mathbb{P}^{2}$ defined by $(x,y,1) =
 (\wp_{\Lambda_{\tau}}(z),\wp'_{\Lambda_{\tau}}(z),1)$ is well defined since 
$\wp_{\Lambda_{\tau}}$ is $\Lambda_{\tau}$-periodic. 
In particular, the elliptic curve $\mathcal{E}_{\tau}$ has $N$\emph{-torsion points} that satisfy 
$N z = 0 \mod \Lambda_{\tau}$. These points correspond to rational points 
$\rational/\integers \times \rational/\integers\tau$ on $\mathcal{E}_{\tau}$, and they are the elements
of the torsion group $\mathcal{E}_{\tau}[N]$.

We will only need to consider 2-torsion points, i.e., points that satisfy 
$2z = 0 \mod \Lambda_{\tau}$ or $z=-z \mod \Lambda_{\tau}$, 
which have a particularly simple description. 
Clearly the non-trivial 2-torsion points for $\complex/\Lambda_{\tau}$ are 
\begin{equation*}
z_1 = \frac{1}{2}\;, \quad z_2 = \frac{\tau}{2}\;, \quad  z_3 = \frac{1 + \tau}{2} \;,
\end{equation*}
which are the crossed points in Fig.\,\ref{fig:LN1_Fig1}\,(a).
Since $\wp(-z) = \wp(z)$ and $\wp'(-z) = -\wp'(z)$, the 2-torsion points $P = -P$ in 
$\complex\mathbb{P}^{2}$ satisfy 
\begin{equation*}
P = (x, y,1) = -P = (x,-y,1), \quad P\in \mathcal{E}_{\tau}.
\end{equation*}
The 2-torsion points are therefore given by the roots of the planar cubic $\mathcal{E}_\tau$,
\begin{equation}
y^2 = 4x^3 + Ax + B = 4 (x - e_1) (x - e_2) (x - e_3)\;,
\label{eq:roots}
\end{equation}
and $(\wp(z_{i}),\, \wp'(z_{i}) ) = (e_i,\, 0)$, with $e_1 + e_2 + e_3 = 0$ 
since the quadratic term vanishes.  The roots are distinct iff 
\begin{equation*}
\Delta = -(A^{3} + 27B^2) = 16(e_{1}-e_{2})^{2}(e_{1}-e_{3})^{2}(e_{2}-e_{3})^{2} \neq 0 .
\end{equation*}

Consider now, for the sake of generality, how the isomorphisms $\mathcal{E}_{\tau} \to \mathcal{E}_{\gamma(\tau)}$ 
induced by transformations $\gamma\in\Gamma$ act on $N$-torsion points. 
We have seen that for any $\gamma\in\Gamma(1)$, the isomorphism acts as
\begin{equation*}
f_{\gamma}: \complex/(\integers+\integers\tau) \mapsto 
\complex/(\integers+\integers\gamma\cdot \tau), \;\;z\mapsto \f{z}{c\tau + d}.
\end{equation*}
We discuss the level $N$ groups  $\Gamma_{0}(N)$, $\Gamma^{0}(N)$, $\Gamma_{\theta}(N)$\;, 
 and $\Gamma(N)$ in turn, following Ref.\,\onlinecite{Silverman}.

Consider first how $\gamma\in\Gamma_{0}(N)$, which has $c = 0 \mod N$ and 
$a=d=1 \mod N$, acts on the cyclic subgroup of order $N$ given by 
\begin{equation*}
C_{N} = \left\{\f{1}{N},\dots, \f{N-1}{N}\right\}
\subset \complex/(\integers+\integers\tau) \;,
\end{equation*}
with $\tau\in\halfspace/\Gamma_{0}(N)$.  We see that $1/N \mapsto 1/N(c\tau+d)$, 
and it follows that 
\begin{eqnarray*}
\f{k}{N} - \f{k}{N(c\tau+d)} &=& \frac{k}{N}  \frac{c\tau + d - 1}{c\tau+d}  \\
\in f_{\gamma}(\integers+\integers \tau) &=& 
\integers+\integers\gamma\cdot\tau = 0\;,
\end{eqnarray*}
for any $k/N \in C_{N}$ and $k=1\,, \dots\,, N-1$. Thus the points $k/N$ and $k/N(c\tau+d)$ 
are equivalent mod $\Lambda_{\gamma\cdot \tau}$ and map to each other in 
$\complex/(\integers+\integers \gamma\cdot\tau)$.

We have seen that $\halfspace/\Gamma(1)$ parametrizes equivalence classes of 
elliptic curves that are isomorphic under an admissible coordinate transformation 
that need only preserve one point, namely the origin which
``rigidifies" a torus to an elliptic curve.   
Similarly, $\halfspace/\Gamma_{0}(N)$ is a moduli space for the equivalence classes 
of $(\mathcal{E}_{\tau},P_{N})$, where $P_{N}$ is a point of order $N$. 
Any point in $C_{N}$ is fixed and $P_{N}$ generates this group. 
In other words, $\halfspace/\Gamma_{0}(N)$ parametrizes elliptic curves up to 
coordinate and modular transformations that preserve both the origin and this additional 
torsion point of order $N$, further ``rigidifying" what we mean by an elliptic curve.

The group $\Gamma^0(N)$, for which $b=0 \mod N$ and $a=d=1 \mod N$, 
carries the points of the cyclic subgroup 
$C^{\tau}_{N} = \set{\tau/N, \dots, \tau (N-1)/N}$ to themselves. 
The moduli space $\halfspace/\Gamma^0(N)$ labels elliptic curves $(\mathcal{E}_{\tau},P^\tau_N)$,
and $\halfspace/\Gamma^0(N) \neq \halfspace/\Gamma_0(N)$ because the two moduli spaces 
are ``tagged" with different torsion points,   $P^\tau_N \neq P_N$.

The group $\Gamma_{\theta}(N)$ has $b=c=0 \mod N$ and $a=d=1 \mod N$ or 
\emph{vice versa}.  It therefore leaves the points $(1 + \tau)k/N$ fixed for all $N$ 
and $k=1,\dots,N-1$, which is another cyclic subgroup of $\mathcal{E}_{\tau}[N]$\,, 
see Fig.\,\ref{fig:LN1_Fig1}\,(a).

$\Gamma(N)$ has $b=c=0 \mod N$ and $a=d=1 \mod N$, from which it follows that
\begin{eqnarray*}
\f{1}{N} - \f{1}{N(c\tau + d)} &=& \f{c\tau + d - 1}{N(c\tau + d)} \\
\in f_{\gamma}(\integers+\integers \tau) &=& 
\integers+\integers\gamma\cdot\tau = 0
\end{eqnarray*}
and
\begin{eqnarray*}
\f{1}{N} \f{a\tau + b}{c\tau + d} - \f{\tau}{N(c\tau + d)} &=& \f{(a - 1)\tau + b}{N(c\tau + d)}\\
\in f_{\gamma}(\integers+\integers \tau) &=& 
\integers+\integers\gamma\cdot\tau = 0 \;.
\end{eqnarray*}
A point in $\halfspace/\Gamma(N)$ is an elliptic curve $\mathcal{E}_{\tau}$ with a basis $\set{1/N, \tau/N}$ for the $N$--torsion subgroup $\mathcal{E}_{\tau}[N]$. 

For $N=2$ these cyclic subgroups have only one point that remains fixed under the morphisms $\mathcal{E}_{\tau}\to \mathcal{E}_{\gamma\cdot\tau}$ from the index 2 subgroups. The other points in $\mathcal{E}_{\tau}[2]$ are interchanged, 
because torsion points map to torsion points. 
Since the common subgroup $\Gamma(2)$ preserves all the $\mathcal{E}_{\tau}[2]$-points, 
the information about these points is the extra structure on the elliptic curve that distinguishes
the moduli spaces $\halfspace/\Gamma$. We have seen that the 2-torsion points 
correspond to the roots $e_{1}$, $e_{2}$ and $e_{3}$ of the cubic equation for 
$\mathcal{E}_{\tau}$.

\subsection{Legendre curves}

Subjecting the Weierstrass form given in Eq.\,(\ref{eq:roots}) to the
change of variables $y\to y/2$, $\Delta\neq0$ and
\begin{equation*}
x = (e_{2}-e_{1})x'+e_{1} \;,\quad y = (e_{2}-e_{1})^{3/2}y' \;,
\end{equation*}
we get
\begin{equation*}
y^{2} = x(x-1)\left(x-\f{e_{3}-e_{1}}{e_{2}-e_{1}}\right) \;.
\end{equation*}
Any elliptic curve can therefore be given in the so-called \emph{Legendre form}:
\begin{equation}
\mathcal{E}_{\lambda}: y^{2} = x(x-1)(x-\lambda) \;,
\label{eq:Legendre}
\end{equation}
where the $\Gamma(2)$-invariant function 
\begin{equation*}
\lambda = \frac{e_{3} - e_{1}}{e_{2} - e_{1}} \neq 0,1,\infty
\end{equation*}
parametrizes the isomorphism classes of elliptic curves with fixed basis of the 2-torsion group. 
We need  to understand in more detail the relation 
between the curves given in eqs.\,(\ref{eq:Weierstrass}) and (\ref{eq:Legendre}).

As we have already discussed, a level 2 curve carries information about the 2-torsion group $\mathcal{E}_{\tau}[2]$. Modular transformations that are in the coset $\Gamma(1)/\Gamma(2)\cong S_{3}$ permute these torsion points, or equivalently the roots of the cubic polynomial of the curve. 
It is instructive to make this explicit.

Assume that two level $2$ curves $\mathcal{E}_{\lambda}$ and $\mathcal{E}_{\lambda'}$ are isomorphic as elliptic curves, i.e.,  that $j(\mathcal{E}_{\lambda}) = j(\mathcal{E}_{\lambda'})$. Then we know that their Legendre equations are related by a change of variables of the form
\begin{equation*}
x= u^{2}x' +r \;,\quad y = u^{3}y' \;,
\end{equation*}
which gives
\begin{eqnarray*}
y'^{2} &=& x'(x'-1)(x'-\lambda')\\
 &=& \left(x'+\f{r}{u^{2}} \right)\left(x'-\f{1-r}{u^{2}}\right)\left(x'-\f{\lambda-r}{u^{2}}\right).
\end{eqnarray*}
There are six possible solutions for the roots $e_{i}(r,u^{2},\lambda)$ ($i=1,2,3$) in terms of permutations of $0,1,\lambda'$,  and we get
\begin{equation}
\lambda' \in \left\{\lambda\;, 1-\lambda\;,  \f{1}{\lambda}\;,  \f{1}{1-\lambda}\;, 
\f{\lambda}{\lambda-1}\;,  \f{\lambda-1}{\lambda}\right\}\;.
\label{eq:orbit}
\end{equation}

\begin{figure}[t]
\begin{center}
\includegraphics[scale = .6]{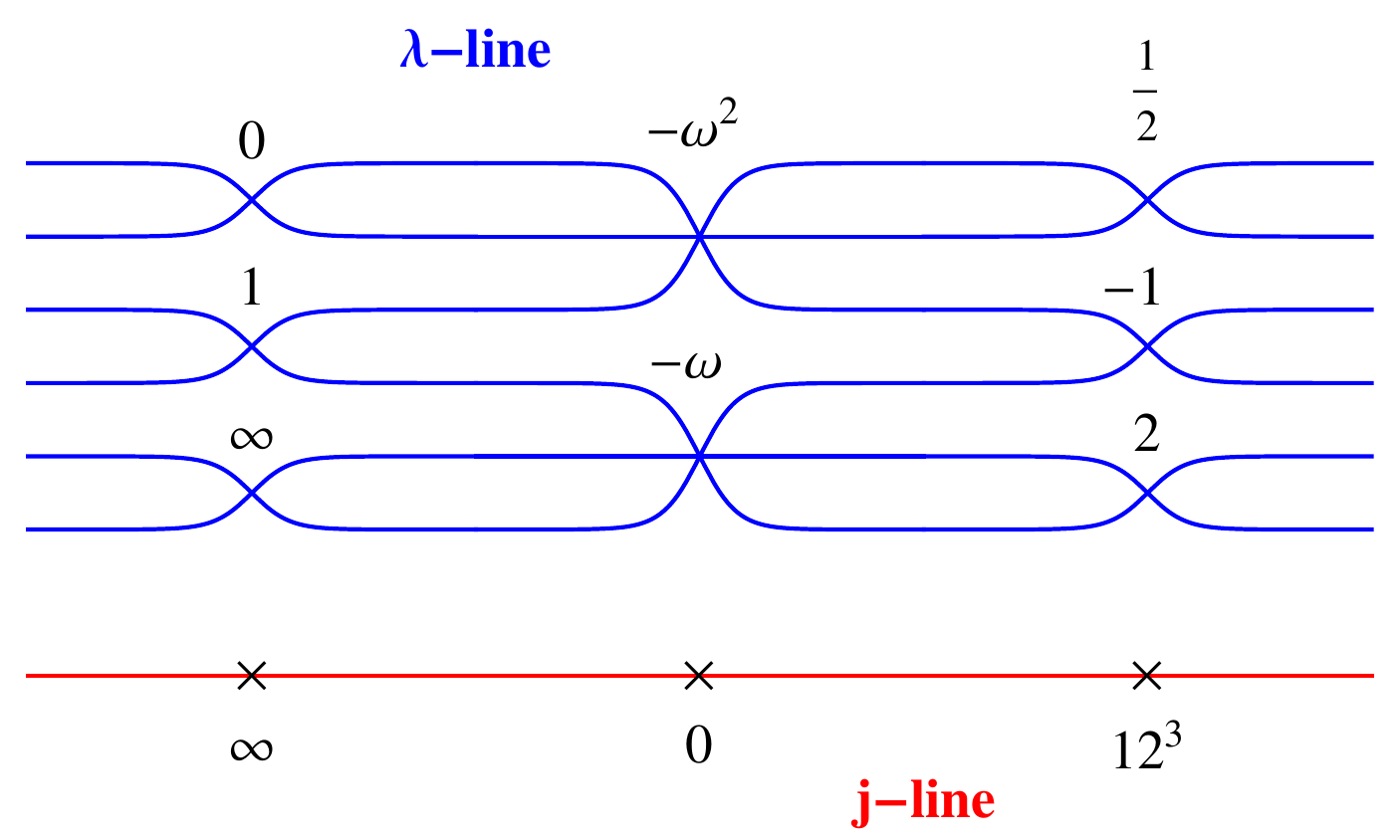}
\end{center}
\caption[j-line]{(Color online) The $\lambda$-line is a six-to-one branched covering over 
the $j$-line.  At branch points ($\omega=e^{2\pi\im/3}$ is a cube root of unity) the curve has extra automorphisms, and the symmetry is enhanced to subgroups between 
$\Gamma(2)$ and $\Gamma(1)$
(adapted from a diagram on p.88 in Ref.\,\onlinecite{DH}).}
\label{fig:LN1_Fig3}
\end{figure}

We see that the parametrization in terms of $\lambda$ is six-to-one in terms of $j$, 
as illustrated in Fig.\,\ref{fig:LN1_Fig3},  corresponding to the index of 
$\Gamma(2)$ in $\Gamma(1)$. 
Indeed, if one writes $j$ in terms of $e_1$, $e_2$ and $e_3$, the expression is invariant under permutations of the roots. Modular transformations in $\Gamma(1)/\Gamma(2)\cong S_{3}$ permute the roots of the equation, and $\lambda =  (e_{3}-e_{1})/(e_{2}-e_{1})$ transforms like the cross ratio of the points $e_{1}, e_{2}, e_{3}$ and $e_{4} = \im\infty$ on the coordinate 
$x\cong \complex\mathbb{P}^{1}$. 
Eq.\,\ref{eq:orbit} is therefore a representation of $S_{3}$ in terms of the modular 
transformations in the coset $\Gamma(1)/\Gamma(2)$. 
 
These properties allow us to write $j$ in terms of $\lambda$.
The change of variables: 
\begin{equation*}
(x,y) \to \left(x + \frac{\lambda + 1}{3}, \frac{y}{2}\right)
\end{equation*}
takes the Legendre curve to the Weierstrass form
\begin{eqnarray*}
y^{2} = 4x^{3} &-& \frac{4}{3}\left(\lambda^{2}-\lambda+1\right) x\\
 &-& \frac{4}{27}(\lambda+1)(\lambda-2)(2\lambda-1)\;,
\end{eqnarray*}
and it follows that 
\begin{equation*}
j(\lambda) = 2^{8}\f{(\lambda^{3}-\lambda+1)^{3}}{\lambda^{2}(\lambda-1)^{2}}\;.
\end{equation*}
This non-singular as long as $\lambda\neq 0,1, \infty$, as it should be.

The function $j(\lambda)$ can be used to show that $j(\lambda) = j(\lambda')$ 
iff $\lambda'$ is on the $S_3$ orbit given by Eq.\,(\ref{eq:orbit}).
It is enough to check this for the two generators of $S_3$, i.e., 
$\lambda\to 1/\lambda$ and $\lambda\to1-\lambda$. 

It follows that the different values $\lambda'$ arise from modular transformations in the coset $\Gamma(1)/\Gamma(2)$. This is useful, since we also know that the coset is of the form
\be
S_{3} \cong \corr{S,T\vert S^{2} = ST^{2}S=  T^{2} = (ST)^{3} = I} \;.
\ee
Combining this with our knowledge of the generators of the subgroups between 
$\Gamma(1)$ and $\Gamma(2)$, we can identify that the order 2 elements in 
Eq.\,(\ref{eq:orbit}) correspond to the elements $S^2$, $T$ and $STS$ 
of order 2 in the coset,
Adding any of these to $\Gamma(2)$ will generate one of the different index 2 subgroups. 
Similarly, the order 3 elements in Eq.\,(\ref{eq:orbit}) correspond to $ST$ and $TS$, 
but if we add one of these we have the other one as well, since $(ST)^2 = TS$ in the coset.
Adding these transformations to $\Gamma(2)$ gives the subgroup $\Gamma_2$ of index 2. 

The above also leads to the observation that fixed points of $\lambda$ under the action \eqref{eq:orbit} correspond to special curves $\mathcal{E}_{\lambda}$ with enhanced modular symmetries generating the subgroups between $\Gamma(1)$ and $\Gamma(2)$. Therefore the $\lambda$-line is no longer 6-to-1 over the $j$-line but less, and the corresponding fixed points can be found in Fig.\,\ref{fig:LN1_Fig3}. 

We have now explained how the $\Gamma(2)$-modular invariant $\lambda$, 
along with its symmetries and relations to the subgroups at level 2, 
arise from the family of level two curves $\mathcal{E}_{\lambda}$.
From these facts it is possible to construct invariant rational functions $f(\lambda)$ 
for the subgroups at level 2.   However, to fix which modular transformation is which in 
Eq.\,(\ref{eq:orbit}), we need to know precisely how the torsion points $e_i$ map to the 
lattice $\complex/\Lambda_{\tau}$. This can be done using theta functions,
and will lead to the most useful explicit representation of $\lambda$.

\section{Modular curves}

In order to find explicit functional forms for the highly transcendental functions discussed
above, we need a better understanding of the level 2 moduli space, i.e., the modular curve
$\mathcal M(2)$ parametrized by ``the $\lambda$-line".
This will give us access to a supply of level 2 modular forms for $\Gamma(2)$ 
and the other level 2 subgroups.  As a by-product we can also
pinpoint the modular transformations in the coset $\Gamma(1)/\Gamma(2)$ 
acting on $\lambda$.

\subsection{Level 2 moduli and theta functions} 

We explicitly construct the branched covering of $\mathcal{E}_{\tau}$ over $\complex\mathbb{P}^{1}$ closely following Ref.\,\onlinecite{Clemens}. 
The functions $\theta_{4}(z,\tau)^{2}$ and $\theta_{1}(z,\tau)^{2}$ on 
$\mathcal{E}_{\tau}$ satisfy
\begin{eqnarray*}
f(z+1,\tau) &=& f(z,\tau) \\
f(z+\tau,\tau) &=& e^{-2\pi\im \tau}e^{-4\pi\im z}f(z,\tau).
\end{eqnarray*}
In other words, they are quasi-doubly-periodic in $z$ with Fourier expansions in $z$. 
It follows that the Fourier coefficient $c_{n+2}$ of $f$ is determined by $c_{n}$. 
The space of holomorphic functions satisfying the above quasi-periodicity is therefore 
at most two dimensional,  and in fact spanned by $\theta_{4}(z,\tau)^{2}$ and 
$\theta_{1}(z,\tau)^{2}$, since they are linearly independent.

There is a map
\begin{equation*}
h: E_{\tau}\cong \complex/\Lambda_{\tau} \to \complex\mathbb{P}^{1}\;, \quad z \mapsto 
\left(\theta_{4}(z,\tau)^{2},\theta_{1}(z,\tau)^{2} \right)\;,
\end{equation*}
which is well defined due to the transformation properties of the theta functions. This is a degree $d=2$ map, so the Riemann-Hurwitz theorem for $h: X\to Y$ gives
\begin{equation*}
2g_{X} - 2 = d \,(2g_{Y}-2) + \sum_{P} (e_{P}-1)\;. 
\end{equation*}
Since $e_{P} \leq \deg h$, the map must be ramified along four points with $e_{P} = 2$. Furthermore,
because $\theta_{4}(\tau/2,\tau) = 0$ and $\theta_{1}(0,\tau) = 0$ are the only zeros, two of the branch points are
\begin{equation*}
h(\tau/2) \sim (0,1)\;, \quad h(0) \sim (1,0)\;.
\end{equation*}
By the lifting property, the following map has a lift from
\begin{equation*}
\complex\mathbb{P}^{1} \to \mathcal{E}_{\tau}, \quad p \mapsto \sum_{P\in h^{-1}(p)} e_{P} P \;.
\end{equation*}
to an entire map on the universal cover $\complex$ of $\mathcal{E}_{\tau}$ (or $\complex\mathbb{P}^{1}$) and therefore by Liouville's theorem is a constant.  
Since $2\cdot 0 = 2 \cdot (\tau/2) =0$ on $\mathcal{E}_{\tau}$, we therefore also have for the other two ramification points $P_{i}$ that
\begin{equation*}
2P_{i}  = 0 \mod \Lambda_{\tau}\;.
\end{equation*}
These branch points are therefore
\begin{eqnarray*}
(\theta_{4}(1/2,\tau)^{2},\theta_{1}(1/2,\tau)^{2})\;\;\\
(\theta_{4}(1/2+\tau/2,\tau)^{2},\theta_{1}(1/2+\tau/2,\tau)^{2})\,.
\end{eqnarray*}
It is convenient to express these theta functions in terms of the Jacobi constants 
$\theta_{i}(0,\tau)$:
\begin{eqnarray*}
\theta_1(1/2,\tau) &=& -\theta_2(0,\tau) \\
\theta_4(1/2,\tau) &=& \theta_3(0,\tau) \\
\theta_4(1/2+\tau/2,\tau) &=& e^{-\pi\im\tau/4}\theta_2(0,\tau)\\
\theta_1(1/2+\tau/2,\tau) &=& e^{-\pi\im\tau/4}\theta_3(0,\tau) \;,
\end{eqnarray*}
which gives the four branch points in the form
\begin{eqnarray}
(1, 0)\; &,& (0,1)\\ 
\left( \frac{\theta_3(0,\tau)^2}{\theta_2(0,\tau)^2},1\right)\; &,&
\left(\frac{\theta_2(0,\tau)^2}{\theta_3(0,\tau)^2},1\right) \;.
\label{eq:thetaleg}
\end{eqnarray}
We see that the branch points are $0$, $1$, $\im\infty$ and 
\begin{equation}
\lambda(\tau) = 
\f{\theta_{2}(\tau)^{4}}{\theta_{3}(\tau)^{4}}\;,
\label{eq:lambdatheta}
\end{equation}
in affine coordinates $x\in\complex\mathbb{P}^{1}$. 
Since the branch points are the images of 2-torsion points of $\mathcal{E}_{\tau}$, 
it follows that the cubic equation of the elliptic curve is given by Eq.\,({\ref{eq:Legendre}), 
and \eqref{eq:lambdatheta} gives an explicit functional form for $\lambda$ whose modular properties are transparent.  We have therefore reconstructed the family of level 2 elliptic curves 
in Legendre form, while obtaining a very useful explicit form for $\lambda$ in terms of 
Jacobi theta-functions.

Observe that our choice of theta functions led to the non-trivial branch point $\lambda$, and other combinations of Jacobi theta function (see the Appendix) give the other values on the
 $S_3$ orbit. These choices also correspond to the permutation of the roots
$e_{1} = 0$,  $e_{2} = 1$ and $e_{3} = \lambda$ of the cubic equation, 
or equivalently the branch points of the cover $\complex\mathbb P^1 \to \mathcal{E}_{\tau}$. 
Since we have four points $0$, $1$, $\lambda$ and $\infty$ on the Riemann sphere 
$\complex\mathbb{P}^{1}$, their cross ratio is an invariant under the automorphisms 
of $\complex\mathbb{P}^{1}$.
We can only reach branch points different from $\lambda$ by permuting 
the points in the cross ratio. This is the action of $S_3$ on $\lambda$.

\subsection{Modular functions at level 2}

Having analyzed elliptic curves and their equations at level 2 in considerable detail, 
we are ready to present the modular functions associated to these moduli spaces.

 A meromorphic function on $\chspace$ that transforms covariantly or contravariantly 
 under the (sub-)modular group is called a \emph{modular function}. 
 These are building blocks of automorphic RG potentials and $\beta$-functions 
 that embody the constraints necessary for any RG flow consistent with the dualities.

For the group $\Gamma(2)$, we have already met an invariant function 
(or a hauptmodul) in Eq.\,(\ref{eq:lambdatheta}).  We have
\begin{eqnarray*}
\lambda(0) &=& 1,\quad \lambda(1)= \infty,\quad \lambda(\im\infty)=0, \\ 
\lambda'(0) &=&  0,\quad \lambda'(1)= \infty,\quad \lambda'(\im\infty)=0.
\end{eqnarray*}
$\lambda$ is an univalent invariant function, so it takes every complex value once in the (proper) fundamental domain of  $\Gamma(2)$ and gives and explicit map $\lambda: \mathbb{H}/\Gamma(2) \to \complex\mathbb{P}^{1}$.

From the transformation properties of the theta functions (Appendix) it follows that
the $\Gamma(2)$-invariant function $\lambda$ transforms in the representation 
\begin{eqnarray}
\lambda\vert_{S} &=&1-\lambda,\quad 
\lambda\vert_{T} = \f{\lambda}{\lambda-1},\quad
\lambda\vert_{STS} = \f{1}{\lambda} \nonumber\\ 
\lambda\vert_{ST} &=& \f{1}{1-\lambda},\quad  
\lambda\vert_{TS} = \f{\lambda-1}{\lambda}
\label{symmLambda}
\end{eqnarray}
under modular transformations in the coset $S_{3}\cong \Gamma(1)/\Gamma(2)$,
and $\lambda\vert_{\gamma}(\tau) := \lambda(\gamma(\tau))$. 
This is the $S_3$ orbit  discussed above (Eq.\,(\ref{eq:orbit})),
and is isomorphic to the one given by Eq.\,(\ref{symm3}); 
$S_{3}\cong\Gamma(1)/\Gamma(2)$ acts as automorphisms \cite{Rankin} of 
$\mathcal{M}(2):=\mathbb{H}/\Gamma(2)\cong \complex\mathbb{P}^1$.  Eq.\,\kaava{symmLambda} gives an explicit formula for the transformation properties of the derivative $\lambda'$, but it is often more convenient to
work with combinations of $\lambda$ and the theta functions (Appendix), since they are more familiar.  

The hauptmoduli for the other groups $\Gamma_{0}(2), \Gamma^{0}(2)$ and $\Gamma_{\theta}(2)$ can all be written as rational functions of $\lambda$:
\begin{eqnarray*}
\lambda_{T} &\equiv \lambda_{0} =& \f{\lambda-1}{\lambda^{2}} \\ 
\lambda_{R} &\equiv \lambda^{0} =& \lambda(1-\lambda)\\ 
\lambda_{S} &\equiv \lambda_{\theta} =& -\f{\lambda}{(1-\lambda)^{2}} \;.
\end{eqnarray*}
It follows from Eq.\,\eqref{symmLambda} that they are invariant under the corresponding groups.

All groups mentioned here have genus-0 modular curves $\mathcal{M}(\Gamma):=\mathbb{H}/\Gamma$, 
which in particular means that there is a one-to-one map $\lambda_{\Gamma}$ to the Riemann sphere $\complex\mathbb{P}^{1}$, see Ref.\,\onlinecite{Rankin}, generalizing the $j$-invariant for $\Gamma(1)$. 
All meromorphic modular forms of weight 0 for $\Gamma$ are rational functions of
$\lambda_{\Gamma}$, and we call this function field 
$M_0(\Gamma) = \complex(\lambda_{\Gamma})$.

Similarly as with invariant functions for the group $\Gamma$, one can define functions that transform covariantly or contravariantly under modular transformations
\begin{equation*}
f(\gamma(\sigma)) = (c\sigma + d)^{k}f(\sigma),\quad 
\gamma\in\Gamma\, , \; \sigma\in\chspace
\end{equation*}
and can be thought of as tensors under modular transfomations. Here the weight $k$ is even, since $-I\in \Gamma$, so that odd forms vanish identically. We also define the notation $f_{k}\vert_{\gamma}:=(c\tau+d)^{-k}f_{k}(\gamma(\tau))$, generalizing the earlier one.

The space of \emph{holomorphic} functions with non-zero positive weight on $\chspace$ is highly constrained, much like the space of invariant meromorphic functions (the weightless case),
 and in fact always finite dimensional. Clearly the space of weight $k$ modular forms $M_{k}(\Gamma)$ for $\Gamma$ is a vector space, whose dimensions \cite{Rankin} for $k = 0$ and $k = 2$ are shown in Table \,\ref{table:modulardim}. 
\begin{table}[h!]
\begin{center}
\scalebox
{1}[1]{
\begin{tabular}{l | c c c c c c}
$\dim M_{k}(\Gamma)$ &$\Gamma(1)$ & $\Gamma_{2}$ & $\Gamma_{0}(2)$ & $\Gamma^{0}(2)$ & $\Gamma_{\theta}(2)$ & $\Gamma(2)$\\
\hline
$k=0$&           1              &          1              &            1                  &             1                &             1                 &           1          \\ 
$k=2$&           0              &          0              &            1                  &             1                &             1                 &           2              \\
\end{tabular}}
\end{center}
\caption{Dimensions of the vector spaces $M_k(\Gamma)$ $(k = 0, 2)$ for level 2 groups.}
\label{table:modulardim}
\end{table}
There is no weight 2 modular form for $\Gamma(1)$. We have already met the weight 4 and 6 forms for $\Gamma(1)$. In fact, these generate all modular forms for $\Gamma(1)$ \cite{Rankin}.

We construct the generators of weight 2 forms for the subgroups that are relevant to the QHE. Since $\lambda$ is a ratio of two theta functions to the fourth power, one can suspect that they have something to do with modular forms for $\Gamma(2)$. In fact, using the transformation properties of the theta functions, one see that
\be
\theta_{2}(\tau)^{4}, \quad \theta_{3}(\tau)^{4}, \quad \theta_{4}(\tau)^{4}
\ee
are all weight 2 forms for $\Gamma(2)$. This space is only two-dimensional and the linear relation between them is the Jacobi identity $\theta_{3}^{4} = \theta_{2}^{4}+\theta_{4}^{4}$.

Since the forms above $\Gamma(2)$ are contained in $M_{2}(\Gamma(2))$, these will be linear combinations of the generators $\theta_{2}^{4}$ and $\theta_{3}^{4}$. 
Inspection shows that 
\bea
\beta^{0}(\tau) &=& \theta_{3}^{4} + \theta_{2}^{4}\\
\beta_{0}(\tau) &=& \theta_{3}^{4} - \f{\theta_{2}^{4}}{2}\\
\beta_{\theta}(\tau) &=& \theta_{3}^{4} -2\theta_{2}^{4}
\eea
are covariant vector fields for the index 3 subgroups. Modular transformations in $\Gamma(1)/\Gamma(2)$ ``permute'' these forms much like for the $\lambda$. Explicitly, $\beta^{0}\vert_{T} = \beta_{\theta}$, $\beta_{\theta}\vert_{T}=\beta^{0}$,
$\beta^{0}\vert_{S} = -2\beta_{0}$ and $\beta_{0}\vert_{S} = -\beta^{0}/2$; but it follows 
that $\beta_{\theta}+\beta^{0} - 2\beta_{0} = 0$, which is necessary since no weight 2 form exists for $\Gamma(1)$. We have now found all the generators for weightless functions 
and weight 2 forms for level 2 symmetries. As shown in Ref.\,\onlinecite{LN2}, all these arise from a family of RG potentials $\Phi_{a}$ as $\beta_{a} = -\doo_{\sigma}\Phi_{a} \propto \theta_{3}^{4}-a\theta_{2}^{4}$.  

 In general, the construction of modular forms may seems to be a daunting task, 
 but the following two results \cite{Rankin} greatly facilitate the job.
 First, let $\lambda_{\Gamma}$ be a univalent\cite{Rankin} weightless modular form 
 for a genus-0 subgroup $\Gamma\subset\Gamma(1)$  
 (all cases considered here are univalent
since the functions have first order pole at one of the inequivalent cusps).
If $\lambda_{q}$ is a modular form with a $q$th order pole at 
$\mathbb{F} - \mathbb{E}$, then
$\lambda_{q} = P(\lambda_{\Gamma})/Q(\lambda_{\Gamma})$ and
\begin{equation*}
\beta_{2k}(\lambda_{\Gamma}) = 
\frac{R(\lambda_{\Gamma})}{S(\lambda_{\Gamma})}\lambda'^{k}_{\Gamma}
\end{equation*}
is a weight $2k$ modular form. In other words, $\beta_{2k}\in M_{2k}(\Gamma)$, 
where $P$, $Q$, $R$, and $S$ are polynomials and the degrees of $P$ and 
$Q$ do not exceed $q$. In particular, the (logarithmic) derivative of a 
weightless modular form is a modular form of weight 2. 
Note, however, that there may be modular forms 
that are not derivatives of modular functions.

Second, let $f$ be a modular form of weight $2k$ for the group
 $\Gamma\subset \Gamma(1)$ of index $\mu$. Then the ``sum rule"
\begin{equation*}
n_{\infty} +\sum_{z_{c} = \gamma(\im\infty), \gamma\notin\Gamma}n_{c} + \sum_{\otimes}\f{1}{\abs{\Gamma_{z_{\otimes}}}}n_{\otimes}+\sum_{p}n_{p} = \f{\mu k}{6}
\end{equation*}
follows from an application of Cauchy's theorem to the fundamental domain.
Here $n_{P}$ is the leading power of $f(z)\sim a_{p}(z-z_{p})^{n_{p}} +\cdots$ around a pole or zero $z_{p}$,  $n_{\otimes}$ is the same quantity for a fixed point 
$z_{\otimes}\in \mathbb{E}$ of $\Gamma$,  $\abs{\Gamma_{z_{\otimes}}}$ 
is the order of the isotropy group of $z_{\otimes}$, and $n_{\infty}$ is the leading power 
of the $q$-expansion of $f(z)$ around the cusp $z=\im\infty$.  

For a level $N$ form $f(\gamma(z)) = (cz+d)^{2k}f(z)$ $\gamma\in\Gamma(N)$, 
there are several inequivalent cusps. The order $n_{c}$ of $f$ at a cusp 
$z_{c} = \gamma_0(\im\infty)$ is defined as the leading order of the 
$q_{N} = e^{2\pi\im z/N}$ expansion of 
$f\vert_{\gamma_{0}}(z)$ at $z=\im \infty$. 
For example, at level 2  $n_{0}$ is the leading power of $z^{-2k}f(-1/z)$ expanded
around $z=\im\infty$ in powers of $q^{1/2}(-z)=e^{-\im\pi z}$.

For the subgroups of index 3 or 6 this reduces to
\begin{equation*}
n_{\infty} + n_{0} + \f{1}{2}n_{\otimes}+\sum_{p}n_{p} = \f{k}{2}\;,
\end{equation*}
or equivalently
\begin{equation*}
n_{\infty} + n_{0}+n_{1} +\sum_{p}n_{p} = k\;. 
\label{order}
\end{equation*}
These theorems are useful since they directly constrain the spaces of modular forms and functions based on the analytic structure. With a bit more work, upper bounds on the dimensions \cite{Rankin} of the relevant spaces $M_{2k}$ cited above can be constructed --- in particular, they are finite.

As explained at length in Ref.\,\onlinecite{LN2} for the QHE, 
this sum rule classifies the fixed points (zeros) or instabilities (poles) of the potentials 
and $\beta$-functions, which correspond to the fixed points and singularities of the 
effective field theories with the given symmetry.

Using the defining $q$-series and transformation properties of the theta functions, we can see that
\begin{eqnarray*}
&\theta_{2}^{4} \quad \theta_{3}^{4} \quad \theta_{4}^{4}\phantom{\,\;.}\\\
n_\infty& 1\quad 0 \quad 0\\
n_1& 0\quad 1\quad 0\\
n_0& 0\quad 0 \quad 1
\end{eqnarray*}
In particular, a linear combination $\beta_{2} \sim a\theta_{2}^{4}+b\theta_{3}^{4}\in M_{2}(\Gamma(2))$ will have a simple zero somewhere on the upper halfplane $\halfspace$.

\subsection{Effective field theories} \label{sec:effective}

The following discussion applies to any quantum field theory with modular symmetry,
but we now start to focus on RG fixed points and critical points of the quantum Hall system.
These points belong to moduli spaces that are  modular curves 
associated with the elliptic curves considered above. 
More details can find be found in the Appendix,  and a comprehensive
account of the symmetries and their representations is given in our 
companion paper\cite{LN2}.

Phenomenology indicates that a modular group $\Gamma$ acts as dualities in the parameter space of the QHE and the quantum critical and fixed points are special points of the modular curve $\mathcal{M}(\Gamma) = \chspace/\Gamma$, 
where the compactification $\chspace$ of $\halfspace$ is the 
physical parameter space.  $\mathcal{M}(\Gamma)$ parametrizes 
elliptic curves with symmetry $\Gamma$, and since any $\Gamma\subset\Gamma(1)$ 
acts properly discontinuously on $\chspace$ it is a Riemann surface. 
For a discussion suitable for our application, see for example Ref.\,\onlinecite{DS}.

The precise form of the automorphism group $\Gamma$ has fundamental 
consequences for the global phase diagram of the QHE. 
Since the quantum Hall plateaux, i.e., the attractive (real) fixed points ($\oplus$), 
are always observed at rational values of the Hall conductance 
$\sigma = \sigma_{H} = p/q \in \rational$, 
they are related to number-theoretic properties of the modular group. 
Filling fractions where no plateau is observed are repulsive rational fixed points 
($\ominus$) in this model.  
All of these follow from the repulsive ``metallic" phase at $\sigma = \im\infty$, 
the filled Landau level $\sigma=1$,
and the attractive quantum Hall insulator at $\sigma=0$ by modular transformations, 
and the set of attractive and repulsive fixed points make up the boundary
 $\doo\chspace = \rational \cup \set{\im\infty}$ 
of the compactified upper half space $\chspace$. 
This follows since the fixed point set $\set{\oplus}\cup\set{\ominus}$ 
(see Fig.\,\ref{fig:LN1_Fig1}\,(d)) maps to itself under modular transformations
\begin{equation*}
\gamma(\infty) = \f{a}{c}\;, \quad \gamma(0) = \f{b}{d}\;, \quad 
\gamma(1) = \f{a+b}{c+d} \;,
\label{cuspimages}
\end{equation*}
and the plateaux-structure in the QHE follows and implies modular symmetries, 
as detailed in the Appendix.

To these boundary fixed points we must add possible RG fixed points inside $\halfspace$.
If we assume that fixed points of $\Gamma$ are the only RG fixed points, then the set of inequivalent RG fixed points is given by the set inequivalent fixed points of $\Gamma$. 
These have the following classification for the full modular group $\Gamma(1)$. \cite{Rankin}

\begin{itemize}
\item[1.] Elliptic fixed points of order 2 ($\otimes$).  
Since these points are fixed by a transformation that squares to one 
it must be conjugate to $S$,
and the set of elliptic points of order 2 is
\begin{equation*}
\mathbb{E}_{2} = \set{\sigma \in \chspace\vert \sigma = 
\gamma^{-1}(\im)\;,\; \gamma\in \Gamma(1)}\;.
\end{equation*}
The isotropy group of any point in $\mathbb E_2$ is $\integers_{2}$.

\item[2.] Elliptic fixed points of order 3.  
Since these points are fixed by a transformation that cubes to one 
it must be conjugate to  $ST$ or $TS$, 
and the set of elliptic points of order 3 is
\begin{equation*}
\mathbb{E}_{3} = \set{\sigma \in \chspace\vert \sigma = 
\gamma^{-1}(\omega)\;,\; \gamma\in \Gamma(1)}\;,
\end{equation*}
with $\omega= e^{2\pi \im/3}$. 
The isotropy group of any point in $\mathbb{E}_{3}$ is $\integers_{3}$. 
We will also use the notation  $\mathbb{E} =\mathbb{E}_{2}\cup \mathbb{E}_{3}$.

\item[3.] Parabolic fixed points or \emph{cusps}. Since these points are fixed by a 
transformation conjugate to $T^n$ ($n\in\integers$), the set of parabolic fixed points is
\begin{equation*}
\mathbb{P} = \set{\sigma \in \chspace\vert \sigma = 
\gamma^{-1}(\infty) \;,\; \gamma\in \Gamma(1)}\;.
\end{equation*}
\end{itemize}
There are also hyperbolic fixed points that are fixed by transformations conjugate to transformations of infinite order, but this is the set of irrational numbers \cite{Rankin} and does not concern us.

The classifications for subgroups $\Gamma\subset \Gamma(1)$ are analogous,
but some fixed point sets may be empty if there are no transformations in $\Gamma$ 
of the type listed above.  In particular, the subgroups at level 2 of index three only have $\mathbb{E}_{2}$ points and the subgroup of index two only points $\mathbb{E}_{3}$, in addition to the cusps. The fixed point sets that do appear are partitioned into 
to equivalence classes mod $\Gamma$.

The interpretation of these fixed points in terms of elliptic curves is the following.
The cusps $\mathbb{P}$ represent degenerate elliptic curves at the boundary of the moduli space, whereas the elliptic points $\mathbb{E}$ are curves with ``extra'' symmetries or automorphisms induced by the fixed-point transformations $\gamma\in\Gamma$. 

Locally, near any fixed point in $\mathbb{E}_{2}$, $\mathbb{E}_{3}$ or $\mathbb{P}$, 
the flow looks the same, since the points are related by an element $\gamma\in \Gamma$ that commutes with the flow. Around a point in $\mathbb{E}_{2}$, the flow bifurcates,
with one irrelevant and one relevant direction. 
Similarly, the flow around a point in $\mathbb{E}_{3}$ ``trifurcates". 

Consider now fundamental domains. By definition, a (proper) fundamental domain 
$\mathbb{F}$ is defined as the set in $\chspace$ so that no two points in $\mathbb{F}$ 
are equivalent under $\Gamma$. We will loosen this definition so that within a 
fundamental domain, two points on the boundary may be equivalent, 
i.e., we take the closure of the proper fundamental domain. 
The topology (fixed point structure) of the RG flow is therefore indeed captured by the modular curve $\mathcal{M}(\Gamma)=\chspace/\Gamma$, 
which is in fact a compact Riemann surface, 
since $\Gamma$ acts properly discontinuously on $\chspace$. 
Considered as a set in $\chspace$, $\chspace/\Gamma$ is the fundamental domain 
$\mathbb{F}_{\Gamma}$ of $\Gamma$. It is easy to see that any fixed point of 
$\Gamma$ is on the boundary of $\mathbb{F}$. \cite{Rankin} 
Therefore, with the assumptions made, the set of critical points is on the boundary 
of the fundamental domain. Moreover, the flow cannot cross the boundaries of 
the fundamental domain, and therefore the boundary segments of 
$\doo\mathbb{F}$ and their images act as separatrices for the flow. 
These boundary segments are also geodesics of the hyperbolic metric on $\chspace$. The covariant $\beta$-function will be a weight two modular form on $\mathcal{M}(\Gamma)$.

We have constructed \cite{LN2} a family of $\Gamma_{a}$ invariant RG flows, 
with $\Gamma(2) \subseteq \Gamma_{a} \subset \Gamma(1)$ 
parametrized by $a\in\real$, so that the critical point $\otimes$ is on the 
boundary of the fundamental domain $\mathbb{F}_{2}$ of $\Gamma(2)$. 
For a general value of $a$, this critical point $\otimes$ on 
$\mathbb{H}$ is not fixed by any element of $\Gamma_{a}$. 
However, the critical point is inherited from a bigger group 
$\Gamma_X\supset \Gamma(2)$ ($X = S\;, T\;, R$), 
which appear for special values $a_{R}$, $a_{T}$, and $a_{S}$ of $a$, 
and in these cases the critical point on $\mathbb{H}$ is fixed by an 
element in $\Gamma_X$.   Otherwise its location moves continuously along 
the boundaries of the fundamental domain 
of the submodular group $\Gamma(2)$ as the parameter $a\in\real$ is varied. 
The set of attractive $\set{\oplus}$ and repulsive $\set{\ominus}$ points is entirely 
determined by the cusps of the subgroups $\Gamma_{X}$ and this highly constraints 
the topology of the global RG flow.

\section{Modular RG flows in the QHE}

Having assembled all the mathematical tools required, we can now build modular models
that contain the universal data and RG flows for quantum Hall systems. 
This will lead us to a 1-parameter family of curves.  The geometry of this family appears 
to encode almost all available scaling data \cite{LR1, LN2}.

\subsection{Renormalization}

An RG flow is a vector field on the space of those parameters that are relevant at
the chosen energy scale.  The QHE is parametrized by the conductivity
tensor $\sigma^{ij}$, or equivalently its inverse, the resistivity matrix $\rho^{ij} = (\sigma^{-1})^{ij}$.
These transport tensors are non-trivial because the background magnetic field breaks
parity (time-reversal) invariance, whence the off-diagonal Hall coefficient is 
permitted by the generalized Onsager relation.
A conventional Hall bar exhibiting fractional plateaux is usually a layered structure 
that confines all charge transport to a single isotropic layer of 
size $L_x \times L_y$, aligned with a current $I$ in the $x$-direction so that 
$R_{*x} = (L_*/L_y) \rho^{*x}$.  
The Hall resistance $R^{\phantom H}_H = \rho^{\phantom H}_H = \rho^{yx}$ is quantized 
in the fundamental unit  of resistance, $h/e^2$, 
while the dissipative resistance $R^{\phantom H}_D = \rho^{xx}/\square$  
is rescaled by the aspect ratio $\square = L_y/L_x$.  

We choose to label the low-energy parameter space by the complexified resistivity 
$\rho = \rho^{xy} + i \rho^{xx} = -\rho^{\phantom H}_H + i \rho^{\phantom H}_D$,
or equivalently by the complexified conductivity  
$\sigma = \sigma^{xy} + i \sigma^{xx} = \sigma^{\phantom H}_H + i \sigma^{\phantom H}_D$.
These complex coordinates only take values 
in the upper half of the complex plane, 
$\mathbb H = \{\sigma\in\mathbb C\vert \Im\sigma = \sigma^{\phantom H}_D>0\}$,
because the dissipative conductivity (resistivity) is positive.
The reason for not including the real line  in our
definition of the parameter space will soon become clear.

The components of the tangent vectors $\beta^{\sigma}= \beta^1 + i \beta^2$ to the RG flow on the space 
of conductivities are the Gellman-Low $\beta$-functions, which describe the (logarithmic) rate of change of the renormalized couplings with respect to the scale parameter $\Lambda$:  
\begin{equation}
\beta^1 = \frac{d\sigma^{\phantom H}_H}{dt}\;,
\quad \beta^2 = \frac{d\sigma^{\phantom H}_D}{dt}\quad\quad
{\rm with} \;\;t = \ln(\Lambda/\Lambda_0).
\end{equation}
The flow ends at stable fixed points, which are the plateaux observed at rational values 
$\sigma_\oplus = \sigma_H\in\mathbb Q$ of the complex conductivity, 
so these are the only real values that should be included in the physical parameter space.
The quantum critical points $\sigma_\otimes\in \mathbb H$ are saddle points of the flow, 
so the $\beta$-function must vanish at these points, and only at these points.
Expanding the $\beta$-function around a critical point gives universal linear coefficients 
$1/\nu_i$, and the real numbers $\nu_i$ are the critical exponents. 

Careful examination \cite{LR1, LN2} of the fixed point structure of scaling data have revealed that there 
appear to be emergent symmetry groups $\Gamma\subset {\rm Aut}\Gamma(1) = 
{\rm Aut}{\rm PSL}(2,\mathbb Z)$, as discussed in sect.\,\ref{sec:effective}.
A first indication of this is that compactifying the topology of the spaces on which 
the modular groups act recovers the plateaux values needed to complete the physical parameter space.  For the full modular group, for example, the space is 
$\overline{\mathbb H} = {\mathbb H} \cup \mathbb Q\cup \{i\infty\}$,
 but is further partitioned into attractive ($\oplus$) and repulsive ($\ominus$) 
 plateaux along the rationals for the subgroups.

This is a very strong statement, with three immediate and unavoidable implications for phenomenology 
(low-energy data) that at first sight appear far too rigid to be of use in physics.  
However, it is precisely this rigidity that gives $\Gamma$-symmetry teeth, and which allows it to be
easily identified.

First, as discussed in the previous section, any $\Gamma$-symmetry partitions parameter space into \emph{universality classes} that are suitable images of $\mathcal{M}(\Gamma)\cong \overline{\mathbb{H}}/\Gamma$, 
with each phase ``attached" to a unique plateau fixed point on the real line, 
see Fig.\,\ref{fig:LN1_Fig1}\,(d).  This is not like a gauge symmetry,
which identifies gauge equivalent field configurations.  Rather, it collects or classifies effective field theories
into families (universality classes) that are related, but not identified, under modular transformations. 
The fixed point set of each $\Gamma$ cannot be manipulated: all the plateaux and quantum critical points
following from $\Gamma$ must be included.   The quantum critical points are located on the phase boundaries, 
and as soon as one is pinned down the location of all the others is fixed.  

Secondly, since the actions of the renormalization group and the modular group must 
be consistent, $\Gamma$ forces the RG flow into a straight-jacket.  
Not only are the RG fixed points, including the 
quantum critical (saddle) points, mapped into each other, 
so are the $\beta$-functions, and it follows that
the critical exponents are ``super-universal": they are always the same, independent of which 
quantum phase transition is considered.  Furthermore, given an initial value for a flow the geometry  of the entire flow-line is largely fixed by the symmetry.   
The flows observed experimentally in the spin-polarized QHE appear to satisfy this 
prediction as well.

Thirdly, it is easy to see that if the $\beta$-function is a holomorphic (complex analytic)  
function of the complex parameter, then the critical exponents can not have opposite signs, 
i.e., the critical point is not a saddle point, which would be in contradiction with experiment.  
Remarkably, it is a mathematical fact that no modular holomorphic $\beta$-functions exist, 
so this cannot happen.  However, non-holomorphic modular forms are very difficult
to handle, and explicit examples are all but non-existent.  
Fortunately, it may not be necessary to abandon analyticity altogether.
The covariant $\beta$-function $\beta_\sigma = g_{\sigma\bar\sigma}\beta^{\bar\sigma}$ 
can be modular and holomorphic.  
This quarantines non-holomorphicity to the metric $g$, a modular form of weight $(2,2)$.  
It is non-singular at the critical points, and therefore invertible, unless the effective field 
theory breaks down at these points.  We have no reason to believe that it will, and this is verified 
\emph{a posteriori} by the agreement of scaling data with the resulting form for the 
$\beta$-function:
\begin{equation}
\beta^{\rm phys} = \beta^\sigma = g^{\sigma\bar\sigma}\beta_{\bar\sigma} =
g^{\sigma\bar\sigma}\partial_{\bar\sigma}\Phi \;.
\end{equation}
This has the consequence that $\nu_1 = - \nu_2$, a phenomenon
dubbed ``anti-holomorphic scaling" in Ref.\,\onlinecite{LR1}.   
The fact hypothesized in Ref.\,\onlinecite{LR1} and explicitly verified in Ref.\,\onlinecite{BL}, 
that modular covariant $\beta$-functions 
are gradients of a ``holomorphically factorized" RG potential 
(this is a slight abuse of notation, since it is
$\exp\Phi(\sigma,\bar\sigma) = \varphi(\sigma)\bar\varphi(\bar\sigma)$
that is holomorphically factorized):
$$
\Phi(\sigma,\bar\sigma) = \ln\varphi(\sigma) + \ln\bar\varphi(\bar\sigma)
$$
is physically reasonable and therefore lends further support to the idea of 
anti-holomorphic scaling.  It is also a very useful simplification, 
since it lifts the geometric analysis of renormalization
from vector to scalar fields.

\subsection{RG potentials}

Our main goal in Ref.\,\onlinecite{LN2} was to construct an RG potential interpolating between all the level 2 symmetries and compare it with available experimental data. We now briefly recollect it. The potential is a $\Gamma$-modular function $\Phi$ whose gradient is the covariant $\beta$-function.     
$\Phi$ is a real-valued modular function that is finite in the finite part of the upper 
half plane, i.e., what we are calling a physical potential.
To this reasonable list of constraints we add that the potential 
should be holomorphically factorized.  This is certainly the case 
at the conformal fixed points, and agrees with the flows
observed in the spin-polarized case.  Together these conditions 
guarantee the existence of a covariant holomorphic $\beta$-function 
that vanishes at the quantum critical points.

The three congruence subgroups $\Gamma_i\;(i = 1,2,3)$ at level two
admit physical potentials $\varphi_i = \ln f_i\;(i = 1,2,3)$, where the $f_i$ are 
elementary modular functions (essentially ratios of theta-functions)
defined in the Appendix.
Since they all contain $\Gamma(2)$ as their largest subgroup,
we consider the most general maximally symmetric 
($\Gamma(2)$-invariant) form for the RG potential,  i.e., 
$\Phi = \psi + c.c.$, where $\psi$ is the linear superposition of 
elementary potentials:
\begin{equation*}
\psi = c_1 \varphi_1 + c_2 \varphi_2 +c_3 \varphi_3\quad (c_i \in\mathbb C).
\end{equation*}
At first sight this ansatz  looks useless,
since it requires six real parameters.  However, for $\Phi$ to factorize
holomorphically the coefficients $c_i$ must be real, giving
\begin{equation*}
\Phi(\sigma,\bar\sigma) =  \ln\vert f_1\vert^{2c_1} + \ln\vert f_2\vert^{2c_2} + \ln\vert f_3\vert^{2c_3}\;. 
\end{equation*}
Up to an arbitrary normalization factor that only affects flow rates, not their shapes,
this leaves only two real parameters.  Furthermore, since it turns out that all the $f_i$ are simple 
fractions of polynomials in the $\Gamma(2)$-modular function $\lambda$, 
the maximally symmetric family of quantum Hall RG potentials reduces to
\begin{equation*}
\Phi_a \propto \ln \lambda(\lambda - 1)^{a-1} + c.c. \;.  
\end{equation*}
This gives the covariant $\beta$-function
\begin{equation*}
\beta_{a}(\sigma) = - \partial_\sigma\Phi_a 
\propto\theta_{3}^{4}(1-a\lambda) = \theta_{3}^{4}-a\theta_{2}^{4}.
\end{equation*}
which vanishes as it should at the quantum critical points
$\sigma_\otimes(a) = \lambda^{-1}(1/a)$. In particular, we can invert this relation
\be
a(\sigma_{\otimes}) = \f{1}{\lambda(\sigma_{\otimes})}. \label{eq:acurve}
\ee
The $\Gamma(2)$-symmetry of the potential $\Phi_a $ along with the covariant 
$\beta$-function $\beta_{a}$ 
is enhanced when $a$ takes certain integer values:  
to $\Gamma_R$ when $a_R = -1$,
to $\Gamma_T$ when $a_T = 1/2$,
and to $\Gamma_S$ when $a_S=2$. These forms at level 2 were discussed above.

Almost all scaling data appear to be consistent with the RG potential $\Phi_a$
for various values of $a$, not restricted to the enhanced symmetry points $a = a_i$, 
but always real. The interpolating forms $\beta_{a}$ for $a\in\real$ between the level 2 
subgroups are shown in Fig.\,\ref{fig:Fig4}.
\begin{figure}[h!]
\begin{centering}
\includegraphics[width=90 mm]{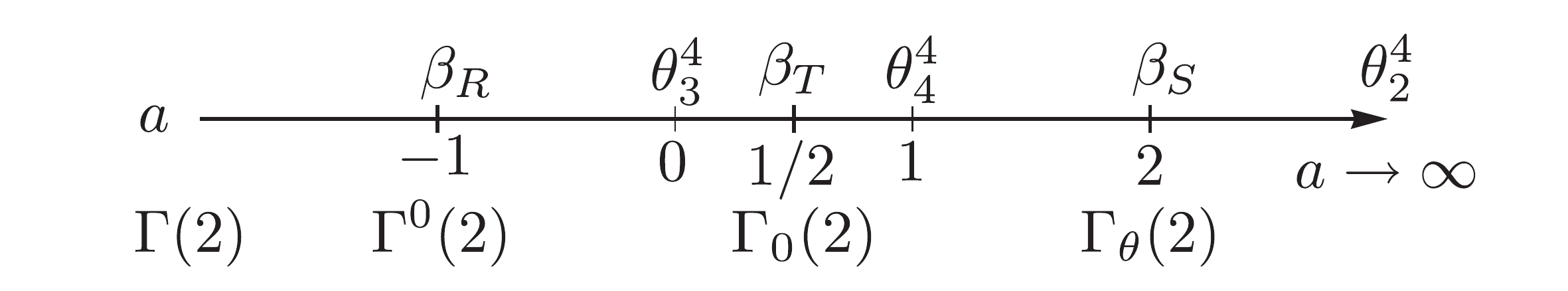}
\caption[summaryofavalues]{The vector field $\beta_a$ is $\Gamma(2)$ symmetric for any value of the parameter $a\in\mathbb C$.
When $a = -1, 1/2, 2$ the symmetry is enhanced to an index 3 group, and when $a = 0, 1, \infty$  
the $\beta$-function degenerates to a $\theta$-function.}
\label{fig:Fig4}
\end{centering}
\end{figure}

\subsection{Quantum Hall curves} \label{sec:Hallcurves}

We can make use of the level 2 curves that we have explained. The relation \eqref{eq:acurve} can be considered to describe as the non-trivial  branch point of level 2 elliptic curve in Legendre form. This immediately allows us to interpret all the universal properties of the QH system exhibiting the scaling with $\Phi_{a}$ as originating from the elliptic curve
\be
\mathcal{E}_{a}: y^2 = x(x-1)(x-a). 
\ee
As long as the value of $a$ is not a fixed point under the $S_{3}$ action on $a= 1/\lambda$, this QH system will have a $\Gamma(2)$-symmetry. For fixed points, the $\Gamma(2)$ symmetry is enhanced to one of the subgroups at level 2. For real values of $a$, only the index 3 groups can appear, and in between the enhanced symmetries the flow satisfies the generalized semi-circle law as explained in Ref.\,\onlinecite{LN2}.

The non-trivial branch point $a$ describes the critical saddle-point $\sigma_{\otimes}$ and the other branch points $0, 1, \im\infty$ correspond to filled Laundau level $\sigma=1$, the QH insulator $\sigma=0$, and the repulsive ``metallic" point $\sigma=\im\infty$, respectively, which are the special points at the boundary of the parameter space.

Finally, the weight 2 modular form $\beta_{a}=-\doo_{\sigma}\Phi_{a} = \theta_{3}^{4}-a\theta_{2}^{4} \in M_{2}(\Gamma(2))$, unique up to real normalization, generates the RG flow of the system, and by construction has a saddle-point zero at $\sigma_{\otimes}(a)$.

As we have shown in Ref.\,\onlinecite{LN2}, changing the value of $a\in \real$ continuously interpolates between the different subgroups, apart from discontinuities at $a=0,1,\im\infty$, where the curve degenerates, which are necessary to exist between the enhanced-symmetry curves at the values $a=-1,1/2,2$,  which all have different sets of attractive ($\oplus$) and repulsive ($\ominus$) plateaux in their moduli.

In summary, the family of RG flows in Ref.\,\onlinecite{LN2} and its phenomenological implications arise from the family of elliptic curves $\mathcal{E}_{a}$.

\subsection{Modular action on the parameter $a$} 

Finally, we show how to deduce the interpretation of the parameter $a$ given above 
without using the location of the critical point $\sigma_{\otimes}(a)$.

The RG potential $\Phi_{a}$ and covariant $\beta_{a}$-function are $\Gamma(2)$ symmetric for any value $a\in\complex$. Under modular transformations in the coset $S_{3}\cong \Gamma(1)/\Gamma(2)$, the family $\set{\Phi_{a},\beta_{a}}$ transforms to itself and we can compute the corresponding modular action on the parameter $a$. 

The generators $S$ and $T$ act on the parameter $a$ via induced actions $S^{*}$ and $T^{*}$ (i.e., as ``pullbacks''), in the sense that for any $\gamma \in \Gamma(1)$, one has
\begin{eqnarray*}
\Phi_{a}(\gamma(\sigma))  &=&\nu(\gamma,a)\; \Phi_{\gamma^{*}a}(\sigma) + \textrm{constant}\\
\beta_{a}(\gamma(\sigma)) &=& f(\sigma,a |\gamma) \; 
\beta_{\gamma^{*} a}(\sigma)\;. 
\label{pullback}
\end{eqnarray*}
One can see that $\beta_{\gamma^{*}a}(\sigma)$ should be a modular form for 
$\gamma^{-1}\Gamma_{a}\gamma$, if $\beta_{a}$ is a form for $\Gamma_{a}$.
Naturally, $\gamma^* a = a$ for any $\gamma\in \Gamma(2)$, 
so in reality the action is non-trivial for $\gamma\in \Gamma(1)/ \Gamma(2)$. 
The action $\gamma^*$ inherits the group law from the action of $\Gamma(1)$ on $\chspace$. 

The multiplier factor $f(z, a | \gamma)$ is the modular weight 
$(\gamma : z)^{2k}:=(cz+d)^{2k}$ for $\gamma\in \Gamma_{a}$. 
Otherwise it differs and in general has an $a$-dependent factor. 
We have the consistency conditions:
\begin{eqnarray*}
\nu(\gamma_1, \gamma_2^*a)  \;  \nu(\gamma_2, a) &=& \nu(\gamma_2\gamma_1, a)\\
f(\gamma_1(\sigma), a \vert\gamma_1) \; f(\sigma, \gamma_2^*a \vert \gamma_1) 
&=& f(\sigma, a \vert\gamma_2\gamma_1)\;.
\end{eqnarray*}

Both the actions of $S^{*},T^{*}$ and the multipliers $\nu(\gamma,a),\ f(\sigma, a |\gamma)$ can be deduced from the known modular properties of $\lambda$ and theta functions.
From
\begin{eqnarray}
\theta_{3}^{4}\left(a\lambda-1\right)\vert_{T} &=& \theta_{3}^{4}\left((1-a)\lambda - 1\right) \\
\theta_{3}^{4}\left(a\lambda-1\right)\vert_{S} &=& 
(a - 1) \; z^{2} \; \theta_{3}^{4}\left( a\lambda/(a-1) - 1\right) 
\label{modulara}
\end{eqnarray}
we get
\begin{eqnarray*}
\nu(T,a) &=& 1,\quad \;\;\;\nu(S,a) = a-1\\
f(\sigma, a \vert T) &=& 1\;,\quad f(\sigma, a\vert S) = (a-1)z^2\;.
\end{eqnarray*}
So we see that $f(\sigma, a \vert \gamma)=\nu(\gamma,a)(\gamma : z)^{2k}$. Then
\begin{equation*}
T^{*}a =1-a, \quad  S^{*}a = \f{a}{a-1} \label{aS3gen}
\end{equation*}
and this specifies the action of the modular group on the parameter $a$.  For the non-trivial transformations $\gamma\in\Gamma(1)/\Gamma(2)$ we obtain
\begin{equation}
a \stackrel{\gamma^{*}}{\longrightarrow} 
\left\{a\;, 1 - a\;,  \f{1}{a}\;, \f{1}{1 - a}\;, \f{a}{a - 1}\;,  \f{a - 1}{a}\right\}\;. \label{eq:aS3}
\end{equation}
These transformations generate a representation of $S_{3}$ isomorphic to the one in Eq.\,\kaava{symmLambda} by the inversion $\lambda\to1/\lambda$. Under modular transformations that are not in the symmetry group, the complexified $\beta$-functions get scaled by an $a$-dependent factor as in Eq.\,\kaava{modulara} and transforms according to the $S_{3}$ transformations in Eq.\,$\kaava{symmLambda}$. This factor can be absorbed into the unknown normalization.

The points of enhanced symmetry correspond to the fixed points of $S_3$ under the group action \eqref{eq:aS3}, since then there is extra modular transformations in the group $\Gamma_{a}$, 
see Fig.\,\ref{fig:LN1_Fig3}. 
The fixed points of the transformations generated by $S^{*}$ and $T^{*}$ are
\begin{equation*}
a = \left\{\pm 1\;, 0\;, 2\;, 1/2\;, -\omega^2, -\omega, \im\infty\right\}\;,
\end{equation*}
see Fig.\,\ref{fig:LN1_Fig3}. 
The real values $a = \set{a_W, a_T, a_S} = \set{-1, 1/2, 2}$ correspond to the groups 
$\Gamma^{0}(2)$, $\Gamma_{0}(2)$ and $\Gamma_{\theta}(2)$ that we have been discussing.
The complex third roots of $-1$ are fixed under the transformations $TS$ and $ST$. 
These extra symmetries generate the group $\Gamma_{2}$. 
In this case, however, the $\beta_{a}$ is not a modular form of weight 2,  
but a modular form with a non-trivial multiplier since the multipliers are complex. 
The multipliers in this case are $\omega$ and $\omega^{2}$, respectively. 
There is no holomorphic weight 2 object for 
$\Gamma_{2}$, so this is the best we can do. 
Note also that $\varphi_{a=1} =\lambda$, $\varphi_{a=0} = \lambda/(\lambda - 1)$ and $\varphi_{a=\im\infty} = \lambda-1$, 
so in these cases, there is no enhancement of symmetry.  Instead we get the canonical 
potential $\lambda$ for $\Gamma(2)$, its translation $\lambda\vert_{T}$ and ``inversion'' $\lambda\vert_{S}$. These lead to the $\Gamma(2)$ forms $\im\pi\theta_{4}^{4}, \im\pi\theta_{3}^{4}$ and $\im\pi\theta_{2}^{4}$, respectively.

\subsection{Singular curves}

As an example of the utility of this algebraic machinery, we analyze the UV/IR limits
of quantum Hall curves, that is, the degenerating behaviour of the curves at the boundary of the moduli space that depends on the family of curves considered.

The Legendre family $\mathcal{E}_{\lambda}$ in Eq.\,(\ref{eq:Legendre}) gives a non-singular elliptic curve 
for every value of the parameter
$\lambda \in  \complex\mathbb P^1  \backslash \{0, 1, \infty\}$.
The curves $\mathcal{E}_0$, $\mathcal{E}_1$ and $\mathcal{E}_\infty$ are singular and represent degenerations 
of the Hall curve at the boundary of moduli space. 
The corresponding complex structures are $\lambda(\tau = 0) = 1$, 
$\lambda(\tau = 1) = \infty$ and $\lambda(\tau = i\infty) = 0$, and clearly give rise to degenerate tori. 

It is useful to write the cubic $\mathcal{E}_\lambda$ in homogenous coordinates 
$$
\mathcal{E}_\lambda: Y^2 Z = X^3 - \frac{\lambda_1}{\lambda_0} X^2 Z - X^2 Z + 
\frac{\lambda_1}{\lambda_0} X Z^2 \;.
$$ 
The singular curves correspond to the values  
$\lambda\simeq (\lambda_0 : \lambda_1)$ as $(1 : 0)$, $(1 : 1)$ and $(0 : 1)$, so
\begin{eqnarray*}
\mathcal{E}_0 : Z Y^2 - X^2 (X - Z) &=& 0 \\ 
\mathcal{E}_1 : Z Y^2 - X (X - Z)^2 &=& 0 \\ 
\mathcal{E}_\infty : \quad\quad\;\;      Z X (X - Z) &=& 0 \;.\\ 
\end{eqnarray*}
$\mathcal{E}_\infty$ is the three concurrent lines in $\complex\mathbb P^2$ 
given by $X = 0$, $Z = 0$ and $X = Z$ that intersect at $(0 : 1 : 0)$. 
In order to analyze $E_1$ we make a change of variables $X \rightarrow X - Z$, giving 
$$
\mathcal{E}_1 : Y^2 Z = X^2 (X + Z )\;.
$$
This is the nodal curve $y^2 = x^3 + x^2$ in affine coordinates, which is 
topologically a sphere with a double point.  This is seen by considering the map 
$\complex\mathbb P^1  \rightarrow \mathcal{E}_1 \subset \complex\mathbb P^2$ given by
$$
 (U : V) \rightarrow \left((V^2 - U^2) U :  (V^2 - U^2)V : U^3\right)\;,
$$ 
which is surjective onto the cubic $\mathcal{E}_1$. 
In the affine patch $Z = 0$ there is only one point 
$(0 : 1 : 0)$ on the curve, which is the point $O$ at infinity.
In the affine patch $Z \neq 0$  $\mathcal{E}_1:  y^2 = x^2 (x + 1)$,  and we set 
$u = Y/X = V/U$.  Then $y = u x$  and $u^2 x^2 = x^2 (x + 1)$ or
$$
x = u^2 - 1, \,\, y = u^2 (u^2 - 1)\;,
$$ 
which is an explicit rational parametrization of the curve on the patch $Z\neq 0$.
However, the map is not injective everywhere, since at the node $(0 : 0 : 1)$ 
(in the original coordinates $(1 : 0 : 1)$)
$$
u = V/U = y/x = \pm 1\;,
$$ 
which confirms that the nodal cubic is topologically a sphere with a double point. 
$\mathcal{E}_0$ is also a nodal curve.  This time the rational map 
$\complex\mathbb P^1 \rightarrow \mathcal{E}_0 \subset \complex\mathbb P^2$ is given by
$$
(U : V) \rightarrow \left((V^2 + U^2) U :  (V^2 + U^2)V : U^3 \right)\;,
$$ 
and the double point is at $(0 : 0 : 1)$ with $u = \pm i$. 

This shows that a quantum Hall curve degenerates to singular geometries at
the cusps, but because these points are infinitely far away from the scaling 
region where the curve accounts for scaling 
data  this does not appear to invalidate the modular model.

\section{Concluding remarks}

We have  shown how the modular symmetries that have been found to be consistent 
with most available scaling data from quantum Hall systems\cite{LR1, LN2} 
derive from a family of algebraic curves  of the elliptic type.  

Holomorphic modular symmetries just barely admits the 
existence of functions and forms that we need in order to give a quantitative 
description of observed RG flows, but no more. Unphysical
holomorphic $\beta$-functions are prohibited by the symmetry.  We have here 
traced this to the geometry of (a family of) underlying algebraic curves that 
encode the experimental data in a neat and compact way.  
The rigidity of holomorphic modular symmetries follows from the rigidity 
of the geometry of elliptic curves, and this is what allows us to harvest 
an infinite number of detailed quantitative predictions.  

The complicated special functions needed to describe scaling data arise
in a transparent way from the group theory and geometry of these quantum Hall curves,
and therefore do not have to be postulated to be the relevant basis for phenomenological
``curve fitting".  The  RG potential emerges naturally in a geometric setting that 
complements the phenomenology found in our companion paper\cite{LN2}.  

We have argued that the algebraic geometry of elliptic curves is an efficient
way to analyze scaling data, extract the modular symmetries of the transport 
coefficients, and use this information to fit the given system into the
one-dimensional (real) family of curves that may model all universal properties 
of quantum Hall systems.  In particular, we have explained how to find the 
quantum Hall curve directly from the most robust part of the scaling data.  
Given such a curve, we have explained 
how the experimentally relevant phase and flow diagrams are obtained from the
geometry of the modular curve that always accompanies an elliptic curve.

Finally, we have discussed the geometry of quantum Hall curves at all RG fixed
points.  This geometry is singular at the cusps (UV and IR fixed points), 
and we have described the non-singular geometry at quantum critical points at length.
Unfortunately the conformal quantum critical model expected to appear
at these points still eludes us.

\section*{Acknowledgement}
CAL is grateful to Jesus College, Oxford for support, and especially to 
Andrew Dancer for patient tutorials on algebraic curves and their degenerations.

\section*{APPENDIX} 

We collect here some facts about modular groups and their 
representations that are used in the main body of this paper. 

\subsection{Modular symmetries, functions and forms}

The modular group $\Gamma(1) := {\rm PSL}(2,\integers)$ 
is generated by $S(\tau) = - 1/\tau$ and $T(\tau) = \tau+1$.  They 
satisfy the algebraic constraints  $S^2 = I$ and $(ST)^3 = (TS)^3 = I$, 
which is the abstract (representation independent) definition of this group.

A {\it modular function} of weight $k$ for $\Gamma \subset \Gamma(1)$ is a meromorphic function of $\tau\in \mathbb H_{\infty} := \mathbb{H} \cup \set{\im\infty}$ with the transformation law
\begin{equation*}
f(\gamma(\tau)) = (c\tau + d )^{k}f(\tau)\;,\;\; 
\gamma(\tau) = \f{a\tau+b}{c\tau + d} \;,
\end{equation*}
under a modular transformation
$$
\gamma = \begin{pmatrix} a& b\\ c& d\end{pmatrix}
\in \Gamma \subsetneq {\rm SL}(2,\real) = \textrm{Isom}(\mathbb H)\;.
$$
We will denote $f_{k}\vert_{\gamma}(\tau):=(c\tau+d)^{-k}f(\gamma\cdot\tau)$.

Since a translation $T^{N}$, for some $N$, is part of every submodular group, every function will have a Fourier expansion in
power of $q_N = e^{2\pi \im \tau/N}$. $q_{N}$ maps $\mathbb{H}$ to the punctured open unit disc and $\mathbb{H}_{\infty}$ to the open unit disc. For simplicity, we consider here the case $q := q_1$, with Fourier expansion
$$
f(\tau) = \sum_{n\in \integers} a_{n}q^n \;,
$$
with $a_{n} = 0$ for $n\ll 0$, for a meromorphic function on $\mathbb{H}_{\infty}$.

Modular forms are {\it holomorphic} functions of weight $k$. The non-positive part $\set{a_{n}}_{n<0}$ is called the polar part and is zero for holomorphic forms. In addition, cusp forms are holomorphic forms that vanish at infinity, i.e., $a_0 = 0$. Suitable analytic structure combined with the transformation law typically restrict $f_{k}$ to be an element in a finite dimensional vector space. 

Similarly, for subgroups with several inequivalent cusps, $f_{k}\vert_{\gamma}(\tau)$ is holomorphic in the neighborhood of any inequivalent cusp  $\gamma\cdot \im\infty$ of $\mathbb{F}$ \cite{DS}. 

Clearly if the coefficients $\set{a_{n}}$ of $f(\tau)$ are real, $f$ satisfies $f(-\bar{\tau})=\bar{f}(\tau)$. Then $f(\tau)$ is real along the imaginary axis. In fact, then it will also be real on the boundary $\doo \overline{\mathbb{F}}$ associated to $\Gamma$, since the boundary of $\overline{\mathbb{F}}$ can be obtained by modular transformations in $\Gamma$. The basic $j$-invariant of $\Gamma(1)$, which classifies elliptic curves, is:
\begin{eqnarray*}
j(\tau) &=& \frac{E_4(\tau)^3}{\Delta(\tau)} \\ 
&=& \frac{1}{q} + 744 + 196684 q + 21493769 q^2  + \dots  \;\;,
\end{eqnarray*}
where
\begin{equation*}
\Delta(\tau) = E_{4}(\tau)^{3} - E_{6}(\tau)^{2} \propto \eta^{24}(\tau) \propto 
\frac{{\lambda'}^6}{\lambda^4(1 - \lambda)^2} 
\end{equation*}
is the weight 12 modular discriminant and
$E_{4}(\tau)$ and $E_{6}(\tau)$ are the weight 4 and 6 Eisenstein functions.
Up to a phase the Dedekind $\eta$-function
\begin{equation*}
\eta(\tau) = q^{1/24}\prod_{n=1}^{\infty}(1-q^{n})
\end{equation*}
transforms with weight $1/2$, i.e.,  $\eta(\tau+1) = e^{\im\pi/12}\eta(\tau)$ 
and $\eta(-1/\tau) = \sqrt{-\im\tau}\eta(\tau)$. 

The $j$-function evaluates  to every complex number exactly once on the interior of the fundamental domain. This follows from the fact that $j(\tau)$ has a simple pole at $\tau=\im\infty$, when $\tau$ is in the compactified fundamental domain $\mathcal{M}(\Gamma(1))\cong \mathbb{CP}^1$.

\subsection{Subgroups at level 2}

The modular group has a lot of interesting subgroups that are studied in number theory. Since the quantum Hall plateaux appear at special rationals $\sigma =\sigma_{H} = p/q\in\rational$, the number-theoretic properties of these groups are of direct relevance. The main congruence subgroup $\Gamma(N)$ at level $N$ is defined as
\begin{equation*}
\Gamma(N) = \bset{\gamma\in \Gamma(1)\bigg\vert \gamma= \begin{matriisi} a & b\\ c& d \end{matriisi} = I \mod N}.
\end{equation*}
The simplest candidate $\Gamma(2) = \langle ST^{2}S,T^{2}\rangle$ has index 6 in $\Gamma(1)$. We are interested in the groups between $\Gamma(1)$ and $\Gamma(2)$. 

The coset $\Gamma(1)/\Gamma(2)$ has 6 representatives
\begin{align}
I \simeq \begin{matriisi}1 & 0\\ 0&1 \end{matriisi},\; 
S\simeq\begin{matriisi} 0 & 1\\ 1 & 0 \end{matriisi}, \; 
T\simeq\begin{matriisi} -1 & 1\\ 0 & 1 \end{matriisi} \\ 
P := ST\simeq\begin{matriisi} 0 & 1\\ -1 & 1 \end{matriisi}, \; 
TS\simeq\begin{matriisi} 1 & -1\\ 1 & 0 \end{matriisi} \\
W:=STS\simeq\begin{matriisi} 1 & 0\\ 1 & -1 \end{matriisi}. 
\label{symm3}
\end{align}
In fact, since $(ST)^{3} = (TS)^{3} = S^{2} = I$ and $T^{2} = I \mod \Gamma(2)$, the coset $\Gamma(1)/\Gamma(2)\cong S_{3}$, the symmetric group of three objects. Furthermore, since $P^{2}=(ST)^{2} = T^{-1}S = TS \mod \Gamma(2)$, there are four groups one can define between $\Gamma(1)$ and $\Gamma(2)$, so that $\Gamma(2)\subset \Gamma$. There is a subgroup of index two (sometimes called $\Gamma_P$):
\begin{align*}
\Gamma_2 = \left\{\gamma\in\Gamma(1)\vert\gamma = I, ST, TS \;({\rm mod}\,2) \right\} = \langle ST,TS \rangle.
\end{align*}
The other subgroups are of index three and all conjugate to each other
\begin{align*}
\Gamma_0(2) &= \left\{\gamma\in \Gamma(1)\vert \gamma = I, T \;
({\rm mod}\,2) \right\} = \langle ST^{2}S,T \rangle\\
\Gamma^0(2) &= \left\{\gamma\in \Gamma(1)\vert \gamma = I, STS \;
({\rm mod}\,2)\right\}  = \langle STS,T^2\rangle\\
\Gamma_{\theta}(2) &= \left\{\gamma\in \Gamma(1)\vert \gamma = I, S \;
({\rm mod}\,2) \right\} = \langle S,T^{2}\rangle.
\end{align*}
The groups are conjugate via $\Gamma_{0}(2)=S\Gamma^{0}(2)S$ and 
$\Gamma^{0}(2)=T^{-1}\Gamma_{\theta}(2)T$.

The way the various subgroups above $\Gamma(2)$ treat the rationals can be deduced from
how they act on on the cusps at $0$, $1$ and $\im\infty$ on their fundamental domain. 
By labelling $0\sim\textrm{e}/\textrm{o}$, 
$1\sim\textrm{o}/\textrm{o}$ and $\im\infty\sim\textrm{o}/\textrm{e}$, these will correspond to the parities of the fractions
$p/q\in\rational$ that are grouped into equivalence classes of the cusps 
$\set{0}$, $\set{1}$ and $\set{\im\infty}$ under the group. 
More concretely, using the transformations of the elements of the group mod 2, 
we can see what parities of $p$ and $q$ are equivalent modulo the group. 
We see that the parities of the matrix entries are constrained as
follows (``o" is the set of odd integers,  and ``e" is the set of even integers): 
\begin{align*}
\Gamma_0(2):&          &c \in \textrm{e};& \;\; a=d \in\textrm{o}&\\
 \Gamma^0(2):&         &b \in\mbox{e}; & \;\;a=d \in\textrm{o}&\\
\Gamma_{\theta}(2):&    &(a,d \in\mbox{e};&\;\;b,c\in\mbox{o})\lor
(a,d \in \mbox{o}; \;\;b,c \in \mbox{e})&\\
\Gamma(2):&          & a,d\in {\rm o};& \;\;b,c\in {\rm e}\;.&
\end{align*}
The modular group $\Gamma(1)$ does not distinguish between the parities of the fractions 
$p/q\in\rational$ (all the cusps $\set{0,1,\im\infty}$ are equivalent), whereas the congruence group $\Gamma(2)$ respects the parities of $p$ and $q$ (all the cusps $\set{0,1,\im\infty}$ are inequivalent). The groups of index 2 partition $\set{0,1,\im\infty}$ into two disjoint subsets. In summary, one has the following disjoint sets of cusps in $\doo\chspace = \rational \cup \set{\im\infty}$, that form equivalence classes under the group:
\begin{align*}
\Gamma_{0}(2):& \quad \set{q \in \textrm{e}} \cup \set{q \in \textrm{o}}\\
\Gamma^{0}(2):& \quad \set{p \in \textrm{e}} \cup \set{p \in \textrm{o}}\\
\Gamma_{\theta}(2):& \quad \set{pq \in \textrm{e}} \cup \set{pq \in \textrm{o}}\\
\Gamma(2):& \quad \set{p \in {\rm o}; q \in {\rm o}} \cup \set{p \in \textrm{e}; q \in \textrm{o}} 
\cup \set{p \in \textrm{o}; q \in \textrm{e}}.
\end{align*}
Here the two sets of inequivalent cusps can be chosen to represent either $\set{\oplus}$ and $\set{\ominus}$. This choice merely inverts the direction of the flow and the corresponding potentials are equivalent up to a sign. 

For the index 2 subgroups, these equivalence classes $\set{\oplus}$ and $\set{\ominus}$ of cusps on 
$\doo\chspace$ are the set of attractive and repulsive RG fixed points. In addition, there are critical 
$\mathbb{E}_{2}$ saddle points $\set{\otimes}$ on $\halfspace$, that are critical points of the 
plateau-plateau transition (see Fig.\,\ref{fig:LN1_Fig1}\,(d)).

For the group $\Gamma(2)$, there are no $\mathbb{E}_{2}$ points, and the critical points $\set{\otimes}$ ``condense'' to the real axis and form ``infinite monkey saddles'' 
$\otimes^{\infty}$ at $0$ or $1$ modulo their modular images. 
Then the set of cusps (and the corresponding rationals) 
$\set{0, 1, \im\infty}\sim\set{{\textrm{e}/\textrm{o}},  \textrm{o}/\textrm{o}, \textrm{o}/\textrm{e}}$ 
correspond to the set of fixed points $\set{\oplus, \ominus, \otimes^{\infty}}$. 

Each holomorphic symmetry $\Gamma\subset\Gamma(1)$  determines a unique 
pair of phase diagrams,
depending on whether the fixed point $\infty = 1/0$ is attractive or repulsive. 
 Given this one bit of information $\Gamma$ 
completely determines the structure of the phase diagram, including the exact location 
of all quantum critical points ($\otimes$), as follows.  

If $\infty$ is an attractive (repulsive) fixed point, then a transition
\begin{equation*}
f = p/q = \oplus \leftarrow\otimes \rightarrow\oplus^\prime = p^\prime/q^\prime = f^\prime>f
\end{equation*}
 exists for:
\begin{description}
\item[$\Gamma_0(2)$] iff\;$q$ and $q^\prime$ are even (odd) and $\delta = 2 (1)$
\item[$\Gamma^0(2)$] iff \;$p$ and $p^\prime$ are odd (even) and $\delta = 1 (2)$
\item[$\Gamma_\theta(2)$] iff \;$pq$ and $p^\prime q^\prime$ are even (odd) and $\delta = 1 (2)$\,,
\end{description}
where $\delta = \det(f^\prime,f) = p^\prime q - p q^\prime$.

The subgroup $\Gamma_{P}=\Gamma_{2}$ does not distinguish between the parity of the rationals and has $\mathbb{E}_{3}$ fixed points at $\omega$ and $-\omega^2$ that physically correspond to trifurcated saddle points. Moreover, $\Gamma_{2}$ has no holomorphic modular forms of weight 2. 

\subsection{Sub-modular functions and forms}

An invariant for the main congruence group $\Gamma(2)$ is 
$$
\lambda(\tau) = \f{\theta_2(\tau)^4}{\theta_3(\tau)^4}  \;,
$$
where the Jacobi theta functions are defined below.
The following $q_{2} =q^{1/2}$ expansions \cite{AS} are useful:
\begin{eqnarray*}
\lambda(\tau) &=& 16 q^{1/2} - 128q +704q^{3/2} -3072q^{2}+ \cdots\\
\frac{16}{\lambda(\tau)} &=& \f{1}{q^{1/2}} + 8 + 20q^{1/2} - 62q^{3/2} + 
216q^{5/2} + \cdots \;, 
\end{eqnarray*}
and $\lambda(0)=1$, $\lambda(\im \infty)=0$, $\lambda(1)=\infty$. 

Invariant functions for the level 2 congruence sub-groups $\Gamma_0(2)$,
 $\Gamma^0(2)$ and $\Gamma_{\theta}(2)$ with index 2 are, respectively:
\begin{eqnarray*}
f_1 &=& f_T(\tau) = \f{\Delta(2\tau)}{\Delta(\tau)} 
\propto \lambda_T = \lambda^{-2}(\lambda-1)\\
f_2 &=& f_R(\tau) = \f{\Delta(\tau/2)}{\Delta(\tau)} 
\propto \lambda_{R} = \lambda(1-\lambda)\\
f_3 &=& f_S(\tau) = \f{\Delta(\tau)^2}{\Delta(2\tau)\Delta(\tau/2)} 
\propto \lambda_S = - \lambda(1 - \lambda)^{-2}.
\end{eqnarray*}
The index 3 subgroup $\Gamma_{2}$ has the invariant function
$$
g(\tau) = \left(\f{\lambda+\omega}{\lambda+\omega^{2}}\right)^{3},
$$
but this group is not relevant for the phenomenology of QHE.

The Jacobi theta functions:
\begin{eqnarray*}
\theta_1(\tau) &=&  2 \sum_{n = 0}^{\infty} (-1)^{n}q^{\f{1}{2} (n+1/2)^2}\\
\theta_2(\tau) &=& \sum_{n = -\infty}^{\infty} q^{\f{1}{2} (n+1/2)^2} \\
\theta_3(\tau) &=& \sum_{n = -\infty}^{\infty} q^{\f{1}{2}n^{2}}\\
\theta_4(\tau) &=& \sum_{n = -\infty}^{\infty} (-1)^n q^{\frac{1}{2} n^2}
\end{eqnarray*}
are weight 1/2 forms that have no zeros in $\mathbb{H}$. \cite{Rankin} Clearly $\theta_{i}(0)=\infty$ or 1,
 and $\theta_{i}(\im\infty)=0$ or 1. The Jacobi identity is
\begin{equation*}
\theta_{3}^{4} = \theta_{4}^{4}+\theta_{2}^{4}.
\end{equation*}
Using Poisson resummation, one can obtain the following transformation properties  
\begin{eqnarray*}
\theta_{2}(T(\tau)) &=& e^{\im \pi/4}\theta_{2}(\tau)\\ 
\theta_{3}(T(\tau)) &=& \theta_{4}(\tau)\\
\theta_{4}(T(\tau)) &=& \theta_{3}(\tau)\\
\theta_{3}(S(\tau)) &=& \sqrt{-\im \tau} \theta_{3}(\tau)\\ 
\theta_{2}(S(\tau)) &=& \sqrt{-\im \tau} \theta_{4}(\tau)\\
\theta_{4}(S(\tau)) &=& \sqrt{-\im \tau} \theta_{2}(\tau)\;. 
\end{eqnarray*}
The so-called theta doubling identities are
\begin{eqnarray*}
2 \theta_{2}(2\tau)\;\theta_{3}(2\tau) &=& \theta^{2}_{2}(\tau) \\ 
2 \theta_{2}(2\tau) &=& \theta^{2}_{3}(\tau)-\theta_{4}^{2}(\tau) \\ 
2 \theta_{3}(2\tau) &=& \theta^{2}_{3}(\tau)-\theta_{4}^{2}(\tau)\\
\theta_{4}^{2}(2\tau) &=& \theta_{3}(\tau)\theta_{4}(\tau)\;. 
\end{eqnarray*}
The modular $\lambda$ transforms under $T$ and $S$ as \cite{Rankin}
$$
\lambda\vert_{T} = \f{\lambda}{\lambda-1}\;,\quad 
\lambda\vert_{S} = 1-\lambda\;, \quad 
\lambda'\vert_{S} = -\lambda'  \;.
$$
The doubling formulas and the relation 
$\Delta(\tau) = \left( \theta_2(\tau)\theta_3(\tau)\theta_4(\tau)/2 \right)^8$ 
imply that:
\begin{equation*}
\frac{\Delta(2\tau)}{\Delta(\tau)} = \frac{\theta^8_2(\tau)}{2^8 \theta^4_3(\tau)\theta^4_4(\tau)} 
= \frac{\lambda^2(\tau)}{2^8 (1 - \lambda(\tau))} 
= - \frac{1}{2^8 \lambda_T(\tau)}.
\end{equation*}
Other useful relations \cite{Rankin} are:
\begin{eqnarray*}
\Delta(2\tau) &=& \eta^{24}(2\tau) = 
\theta_{2}^{16}(\tau)\theta_{3}^{4}(\tau)\theta_{4}^{4}(\tau)/2^{16}\\
\lambda_{T} = -\f{\theta^{8}_{2}}{\theta_{3}^{4}\theta_{4}^{4}}&,&\quad
\lambda_{S} = \f{\theta_{2}^{4}\theta_{4}^{4}}{\theta_{3}^{8}}\;\;,\quad
\lambda_{W}= \f{\theta_{2}^{4}\theta_{3}^{4}}{\theta_{4}^8}\\
f_{W}(\tau) &=& \f{\Delta(\tau)}{\Delta(\tau/2)}  
= \f{1}{2^{4}} \f{\theta^{4}_{2}\theta^{4}_{3}}{\theta_{4}^{8}}\\
f^{-1}_{T}(\tau) &=& 16 f_{W}(2\tau) = 16f_{S}(2\tau+1)\\
f_{S}(\tau) &=& f_{W}(\tau+1)  = f_S(\tau+2)\\
f_{T}(-1/\tau) &=& -f^{-1}_{W}(\tau).
\end{eqnarray*}

The derivative of the modular $\lambda$ is
\begin{equation*}
\lambda'(\tau) = \f{4\im}{\pi} K(k)^{2}(1-\lambda)\lambda,
\end{equation*}
where $K(k)$ is the complete elliptic integral of the first kind
$$
K(k) = \int^{\pi/2}_{0} \f{\de \theta}{\sqrt{1-k^{2} \sin^{2}\theta}}  = \f{\pi}{2}\sum_{n=0}^{\infty} \left[\f{(2n)!}{2^{2n}n!^{2}}\right]^{2}k^{2n}.
$$
Some special values are $K(0)=\pi/2$, $K(1)=\infty$, $K(\im\infty)=K(\infty)=0$.

The canonical variable $k(\tau)  = \lambda(\tau)^{1/2}$ is called the elliptic modulus. 
The theta functions are related to elliptic integrals by
\begin{eqnarray*}
\theta_{2}(\tau) = \sqrt{\f{2kK(k)}{\pi}}\;&,& \quad \theta_{3}(\tau) = \sqrt{\f{2K(k)}{\pi}}\\
\theta_{4}(\tau) &=& \sqrt{\f{2k'K(k)}{\pi}}
\end{eqnarray*}
where $k'^2 = 1 - k^2$ is the complementary modulus. 
Eq.\,(7.2.17) in Ref.\,\onlinecite{Rankin} gives
$$
\f{\lambda'}{1-\lambda} = \im\pi\theta_{2}^{4}\;,\quad 
\f{\lambda'}{\lambda(1-\lambda)}=\im\pi\theta_{3}^{4}\;,\quad 
\f{\lambda'}{\lambda} = \im\pi\theta^4_4 \;.
$$
In particular $K(k) = (\pi/2) \theta^{2}_{3}(\tau)$. 
If we define $K'(k) := K(k')$, then $\exp(\pi\im\tau) = \sqrt{q} = \exp( - \pi K'/K)$ and
$$
\tau = \f{\im K'(k)}{K(k)} \mod \Gamma(2)\;.
$$
The $\Gamma(2)$-invariant function $\lambda$ is essentially the inverse of the elliptic nome $q(m)$ in terms of the parameter $m(q) = \lambda$, and in general elliptic integrals are 
the inverse functions from the elliptic curve to $\mathbb{CP}^1$ 
(see Fig.\,\ref{fig:LN1_Fig2}).


\end{document}